\documentclass[twocolumn]{aastex631}

\usepackage{amsmath}

\begin{document}

\title{Radio Burst Phenomenology of AD Leonis and Associated Signatures of Propagation Effects}

\author[0009-0004-0713-405X]{Jiale Zhang}
\affiliation{School of Earth and Space Sciences, Peking University, Beijing 100871, China}
\affiliation{Key Laboratory of Solar Activity and Space Weather, National Space Science Center, Chinese Academy of Sciences, Beijing 100190, China}
\affiliation{ASTRON, Netherlands Institute for Radio Astronomy, Oude Hoogeveensedĳk 4, Dwingeloo, 7991 PD, the Netherlands}
\affiliation{Kapteyn Astronomical Institute, University of Groningen, P.O. Box 800, 9700 AV, Groningen, the Netherlands}

\author[0000-0002-0872-181X]{Harish K. Vedantham}
\affiliation{ASTRON, Netherlands Institute for Radio Astronomy, Oude Hoogeveensedĳk 4, Dwingeloo, 7991 PD, the Netherlands}
\affiliation{Kapteyn Astronomical Institute, University of Groningen, P.O. Box 800, 9700 AV, Groningen, the Netherlands}

\author[0000-0002-7167-1819]{Joseph R. Callingham}
\affiliation{ASTRON, Netherlands Institute for Radio Astronomy, Oude Hoogeveensedĳk 4, Dwingeloo, 7991 PD, the Netherlands}
\affiliation{Anton Pannekoek Institute for Astronomy, University of Amsterdam, 1098 XH, Amsterdam, the Netherlands}

\author[0000-0002-1369-1758]{Hui Tian}
\affiliation{School of Earth and Space Sciences, Peking University, Beijing 100871, China}
\affiliation{Key Laboratory of Solar Activity and Space Weather, National Space Science Center, Chinese Academy of Sciences, Beijing 100190, China}

\correspondingauthor{Jiale Zhang, Hui Tian}
\email{jialezhang@pku.edu.cn, huitian@pku.edu.cn}



\begin{abstract}

\noindent We present the high-resolution radio dynamic spectra of AD Leonis (AD Leo) between 1.0 and 1.5 GHz taken by the Five-hundred-meter Aperture Spherical radio Telescope (FAST) on Dec. 1st, 2023. Over a 15-minute period, we identify complex, superimposed spectro-temporal structures, including: (1) broadband, second-long modulation lanes with downward frequency drifts, (2) narrowband ($\approx$ 50\,MHz), short-duration S-burst envelopes with upward drifts, and (3) even narrower ($\approx$ 10\,MHz), millisecond-scale S-burst striae within these envelopes. Using the discrete Fourier transform and auto-correlation function, we identify two dominant periodic emission patterns, corresponding to the periodicities of the S-bursts ($\approx0.1\,$s) and the striae ($\approx0.01\,$s). The complex superposition of diverse time-frequency structures poses a challenge to interpreting all the emission variability as intrinsic to the source. We propose that the modulation lanes could be a propagation effect as the radio waves traverse an inhomogeneous, regularly structured plasma region in the AD Leo's magnetosphere. By modelling a plasma screen with sinusoidal phase variation in one dimension, we show that we could qualitatively reconstruct the observed modulation lanes. The origin of the finest structures, the striae, remains unclear. Our work highlights that propagation effects in the stellar magnetosphere can potentially probe kilometre-scale structures in the emission regions and provide novel constraints on density inhomogeneities caused by magnetohydrodynamic waves that are difficult to access by other means.
\end{abstract}

\keywords{	
 --- M dwarf stars (982) --- Radio bursts (1339)	--- 
Interstellar scintillation (855)}


\section{Introduction} \label{sec:intro}

AD Leonis (AD Leo, M3.5V) is one of the most intensively studied radio stars in the solar neighbourhood (4.965\,pc, \cite{2020yCat.1350....0G}). The star has been observed for more than three decades and is known to produce some of the brightest radio bursts (up to 0.5\,Jy) with high circular polarization (up to 100\%)\citep{1990ApJ...353..265B,1997A&A...321..841A,2006ApJ...637.1016O,2008ApJ...674.1078O,2019ApJ...871..214V,2021NatAs...5.1233C,2023ApJ...953...65Z,2024MNRAS.531..919Z,2024A&A...686A..51M} among radio-emitting M-dwarfs. The high brightness temperature ($>10^{14}$ K) and circular polarization degree indicate that the emission is produced by a coherent process, likely electron cyclotron maser (ECM) emission \citep{1985ARA&A..23..169D,2006A&ARv..13..229T}. ECM emission occurs at the local cyclotron frequency ($\nu_{ce}\approx2.80\cdot B\;\mathrm{[MHz]}$) or its low harmonics, providing a direct measurement of the local magnetic field strength at the source region.

One of the most intriguing aspects of AD Leo's radio emission is that it has a variety of rich structures in the time-frequency plane (the dynamic spectrum), spanning from hours to millisecond scales \citep{1990ApJ...353..265B,1997A&A...321..841A,2006ApJ...637.1016O,2008ApJ...674.1078O,2019ApJ...871..214V,2023ApJ...953...65Z}. Millisecond-scale information of the radio signals were first revealed by the high-cadence observations from the Arecibo telescope \citep{1990ApJ...353..265B,1997A&A...321..841A,2006ApJ...637.1016O,2008ApJ...674.1078O}. With the recent progress in sensitivity and spectro-temporal resolution enabled by the Five-hundred-meter Aperture Spherical radio Telescope (FAST, \cite{2006ScChG..49..129N,2020RAA....20...64J}), our understanding of AD Leo's radio emission has been greatly advanced. Narrowband, frequency-drifting bursts at millisecond time scale are now identified as a common feature of AD Leo’s radio emission \citep{2023ApJ...953...65Z,2025A&A...695A..95Z}. {Such bursts are commonly compared to S-bursts observed from Jupiter. Jovian S-bursts represent the best example of fine structures being produced by ECM emission \citep{1996GeoRL..23..125Z,2014A&A...568A..53R,2023NatCo..14.5981M}.} Similar to Jovian S-bursts, the millisecond radio bursts detected from AD\,Leo are often narrowband, highly circularly polarised, and displays parallel frequency drifts and periodicities \citep{2023ApJ...953...65Z}. Using the radio emission from Jupiter as a guide, the frequency drifts of these bursts are interpreted as the relativistic motion of source electrons along the magnetic field lines with a certain magnetic field gradient \citep{2023ApJ...953...65Z,2025A&A...695A..95Z}. Measuring the drift rate of the bursts provides constraints on the energy of radio-emitting electrons and the local magnetic topology. For instance, the ultrafast frequency-drifting radio bursts ($\approx$8 GHz/s) recently found on AD Leo imposed limits on the size of the localized magnetic spots on the surface \citep{doi:10.1126/sciadv.adw6116}.

With the recent upgrade in sensitivity, resolution, and bandwidth of observations conducted by radio telescopes worldwide, a wealth of details in the spectro-temporal structures of the radio bursts from nearby stars have been brought to light (e.g. \cite{2019ApJ...871..214V,2021A&A...648A..13C,2022ApJ...935...99B,2024A&A...682A.170B,2025ApJ...990L..32L}). The progress highlights the need to establish a classification framework of these structures at different time scales (e.g. \cite{2024A&A...682A.170B}), and to reach a broad consensus of their physical interpretations \citep{2025A&A...695A..95Z}. One aspect that has not received much attention so far is the potential propagation effects in the corona, which could modulate the intensity of the emission. The refractive index of the stellar corona varies due to the inhomogeneities in plasma density and magnetic field. As radio waves propagate through such plasma medium with spatially-varying refractive indexes, they are scattered by the local density irregularities and then interfere, which might produce observable time-frequency intensity variations similar to interplanetary scintillation \citep[IPS;][]{1973ARA&A..11....1J} and interstellar scintillation \citep[ISS;][]{1990ARA&A..28..561R}.

A characteristic length scale in scintillation is the Fresnel scale, defined as $r_F=\sqrt{\lambda z/(2\pi)}$, where $\lambda$ is the wavelength of the emission and $z$ is the effective distance of the scattering medium (often modelled as a thin plasma screen). If the screen lies much closer to the source than to the observer, $z$ reduces to the source–screen distance. In the strong scattering regime, where the phase variation of the plasma screen $\zeta$ is greater than 1 rad across the Fresnel scale ($\zeta\gg 1\;$rad), scintillations are generally divided into two regimes, refractive and diffractive scintillations \citep{1996ApJ...460..460S}. 

Refractive scintillation arises from focussing and de-focussing of light rays, leading to slow, broadband emission variations. It occurs when the angular size of the source is smaller than the characteristic patch size of the scattering medium, referred to as the refractive scale $r_{\rm ref}\sim (\zeta/1\;\mathrm {rad})\; r_F$. In contrast, diffractive scintillation is caused by constructive and destructive  interference of waves from multipath propagations. It typically occurs at a much shorter time scale and a narrower bandwidth. The characteristic spatial scale is the diffractive scale $r_{\rm diff}\sim (1\;\mathrm{rad}/\zeta) \;r_F$, corresponding to the transverse scale over which the wave phase remains coherent. The two regimes have been well separated in the observations of ISS (see \cite{1990ARA&A..28..561R} for a review). 

In the corona of AD Leo, for instance, for a hypothetical plasma screen at one stellar radius (0.44 $R_\odot$ as the radius of AD Leo, \cite{2015ApJ...804...64M}) from the emission region, the Fresnel scale at L-band is approximately 3 km. ECM emission is believed to arise from elementary emitters which are very compact in size. For instance, the elementary sources of Jovian S-bursts are estimated to be smaller than 10\,km based on the narrowband nature of the emission (instantaneous bandwidth down to 10 kHz, \cite{2014A&A...568A..53R,2015ApJ...809....4G}). Similar constraints on AD Leo radio bursts suggest the elementary emitters smaller than a few hundred kilometres \citep{2008ApJ...674.1078O,2023ApJ...953...65Z}. These scales imply that, under reasonable coronal turbulence conditions, the emission may undergo refractive scintillation, and in some extreme cases, even diffractive scintillation if the source is sufficiently compact.

Our current study was motivated by FAST observation of AD Leo on Dec. 1st, 2023, which revealed radio bursts with remarkably unusual patterns in the dynamic spectra. In addition to the S-bursts that we have already mentioned, we identified two new spectro-temporal features: broadband, second-long modulation lanes on top of the S-bursts and narrowband, millisecond-scale striae inside the S-bursts. In this work, we will characterize and disentangle these complex structures in the dynamic spectra and explore plausible explanations for some of their characteristics, with a particular focus on the possibility of propagation effects.

\section{Observations and results}

We observed AD Leo on December 1st, 2023, from 20:00 to 24:00 with FAST at L-band (1.0 - 1.5 GHz). We used the same observation mode (pulsar backend with a sampling time of 196.608 $\mu$s and 1024 frequency channels) and the same pipeline for data reduction as in \cite{2023ApJ...953...65Z}. {Multi-beam data was inspected to help exclude narrowband radio frequency interferences (RFIs), as well as sporadic broadband RFIs.} Similar to previous observations on AD Leo \citep{2023ApJ...953...65Z}, a large number of millisecond-scale radio bursts were detected, spanning almost the entire 4-hour observations. In this study, we restrict our attention to a 15\,min time period that shows intense radio bursts with a rich variety of spectral structures. The radio bursts are highly left-hand circularly polarized (we adopted the PSR/IEEE convention \citep{2010PASA...27..104V} for the Stokes V sign), reaching up to 100\%. {The sign of the circular polarization is consistent with the x-mode emission from the more visible magnetic south pole of AD Leo from a terrestrial observer \citep{2019ApJ...871..214V,2023A&A...676A..56B,2025A&A...695A..95Z}}. As we found that the Stokes V dynamic spectra show almost identical patterns with the Stokes I data and are typically subject to less interference from the background, only Stokes V data is presented and analyzed in this paper. 

{We found complex and diverse emission structures on different timescales during the 15\,min observation, ranging from seconds to milliseconds. Three most prominent features are the modulation lanes, S-burst envelopes and S-bursts striae, which are introduced in the following sub-sections.}

\subsection{Modulation lanes}\label{sec:modulation}

\begin{figure*}[htbp]
    \centering
    \includegraphics[width=0.95\linewidth]{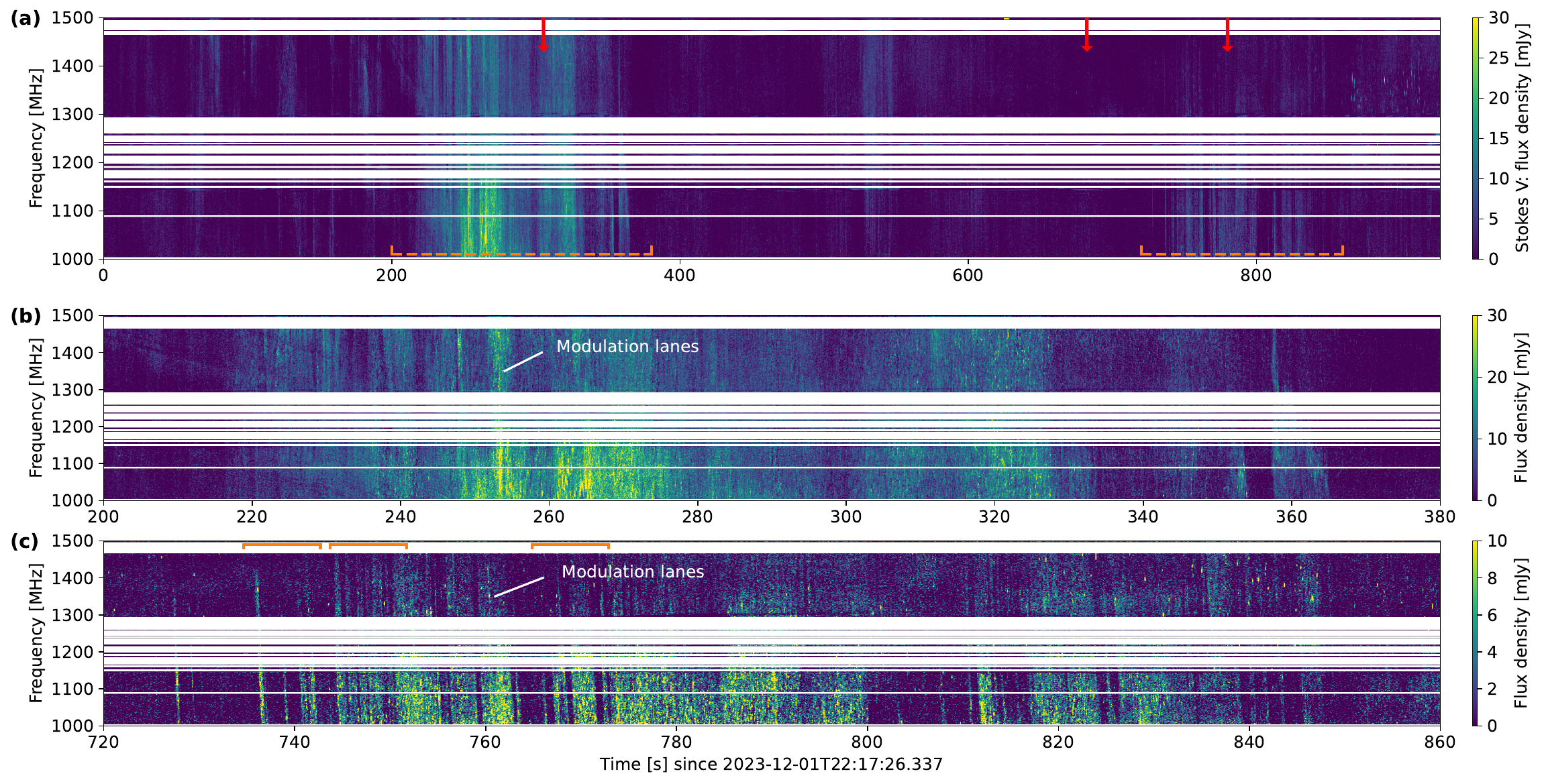}
    \caption{(a) Stokes V radio dynamic spectrum of the 15-minute observation analyzed in this work. Positive values represent left-hand circular polarized emission. The red arrows mark the times of the dynamic spectra in Figure \ref{fig:figure3} (a, c, e) and the secondary spectra in Figure \ref{fig:figure5} (a, d, g). The orange dashed bars mark the time spans of two major clusters of bursts. (b, c) Zoomed-in dynamic spectra of the two clusters of bursts. The orange bars in panel (c) mark the time spans of the dynamic spectra in Figure \ref{fig:figure2} (a, c, e). The blank horizontal gaps represent the corrupted channels with strong RFIs. {Examples of the modulation lanes are indicated in panel (b, c).}}
    \label{fig:figure1}
\end{figure*}

\begin{figure*}[htbp]
    \centering
    \includegraphics[width=0.95\linewidth]{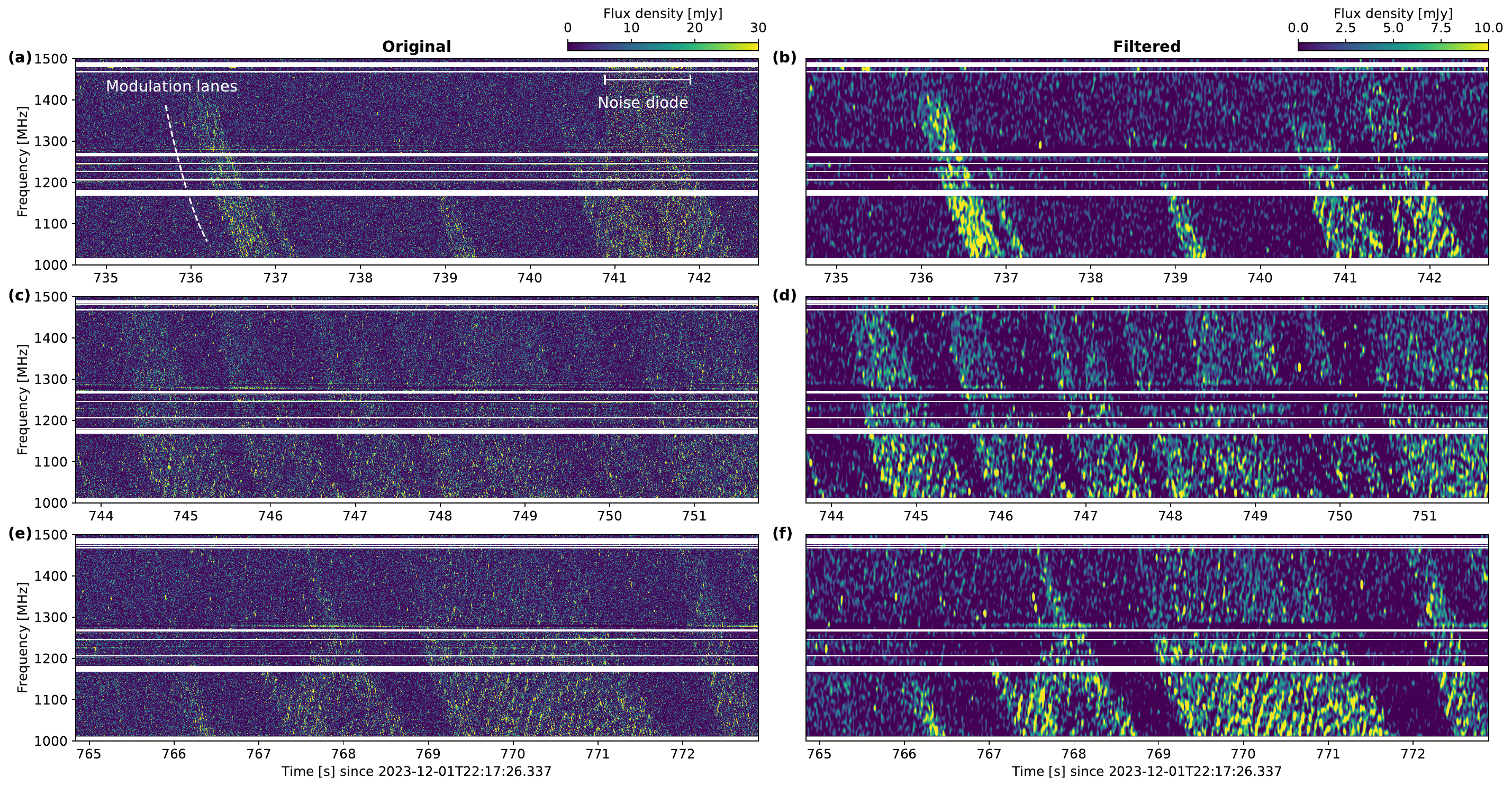}
    \caption{(a, c, e) Examples of the modulation lanes in the dynamic spectra. The time and frequency resolutions of the dynamic spectra are $0.39\;\mathrm{ms}$ and $0.49\;\mathrm{MHz}$, respectively. {The rectangular diffuse pattern marked by the white horizontal line in panel (a) is an instrumental artifact caused by the 1\;s long noise diode signal used for calibration, which overlaps in time with two sweeping modulation lanes.} The overall frequency sweep of the modulation lanes is indicated by a white dashed line. (b, d, f) Corresponding smoothed dynamic spectra after applying a low-pass filter. The low-pass filter smooths out structures smaller than 0.03 s in time or 20 MHz in frequency. Details of the low-pass filter are provided in Appendix \ref{sec:low-pass}. {In panel (b), the artifact related to the noise diode signal is removed after the filter.}}
    \label{fig:figure2}
\end{figure*}

In Figure \ref{fig:figure1}(a), almost vertical burst structures are present in the dynamic spectrum. A closer look at the two clusters of bursts (200 - 380\,s in Figure \ref{fig:figure1}(b) and 720 - 860\,s in Figure \ref{fig:figure1}(c)) reveals broadband emission patterns as lanes which last on time scales of seconds. {In Figure \ref{fig:figure1}(b), a few bright and diffuse lanes can be seen around 260 s, while in Figure \ref{fig:figure1}(c), lanes are identified with clearer edges. We call these emission lanes as the modulation lanes.}

{Figure \ref{fig:figure2} shows zoomed-in examples of these modulation lanes, with the original dynamic spectra on the left as well as smoothed ones (applying a low-pass filter, details are given in Appendix \ref{sec:low-pass}) with better contrast on the right. We found that these modulation lanes cover the entire observing bandwidth and have a typical single-frequency duration of 0.5 - 2\,s.} They share a very similar sweep towards lower frequencies (drift rate of $\approx 0.8 $ GHz/s, slightly faster at higher frequencies). Many of these lanes seem to be thinner at high frequencies (e.g. {lane} at 745 s), meanwhile the gaps in between these lanes seem to be wider at high frequencies. Their appearance could be stochastic (e.g. Figure \ref{fig:figure2} (a, b, e, f)) or quasi-periodic (around 1 s, lanes from 744 s to 749 s in Figure \ref{fig:figure2}(c, d)). The average intensity within the modulation lanes is about 5 mJy and falls less than 1 mJy at the gaps in between. 

{Similar emission patterns have been reported on other M dwarfs \citep{1990ApJ...353..265B} (referred to in the paper as quasi-periodic pulsations) and magnetic massive star \citep{2025ApJ...989..163D} (referred to as spikes), yet no unified terminology or underlying physical mechanism has been established to describe them. In the context of Jovian studies,} these modulating patterns are known as the ``modulation lanes'', typically describing a set of parallel sloping lanes with alternating intensity maximum and minimum seen in the dynamic spectra \citep{1970A&A.....4..180R,1978Ap&SS..56..503R,1997JGR...102.7127I,2009A&A...493..651L,2013Icar..226.1214A}. In most cases, Jovian modulation lanes are seen superimposed on other Jovian fine structures, like Jovian S-bursts or L-bursts. Their origin was suggested to be a propagation effect, possibly due to the refraction from density inhomogeneities of a certain plasma region near the Io's orbit \citep{1997JGR...102.7127I}. 

The modulation lanes in our study are very similar to the Jovian ones in general morphology (e.g. parallel sloping patterns), despite the differences in the emission frequency, the drift rate, and the bandwidth. Another reason that draws us to Jovian analogy here, as we will explain later, is the presence of structures that are much finer than the modulation lanes and appear as some independent emission patterns. This is suggestive of a possible propagation-related modulation in addition to the intrinsic emission variability in the dynamic spectra.

\subsection{S-burst envelopes and striae}\label{sec:fine}

\begin{figure*}[htb!]
    \centering
    \includegraphics[width=0.95\linewidth]{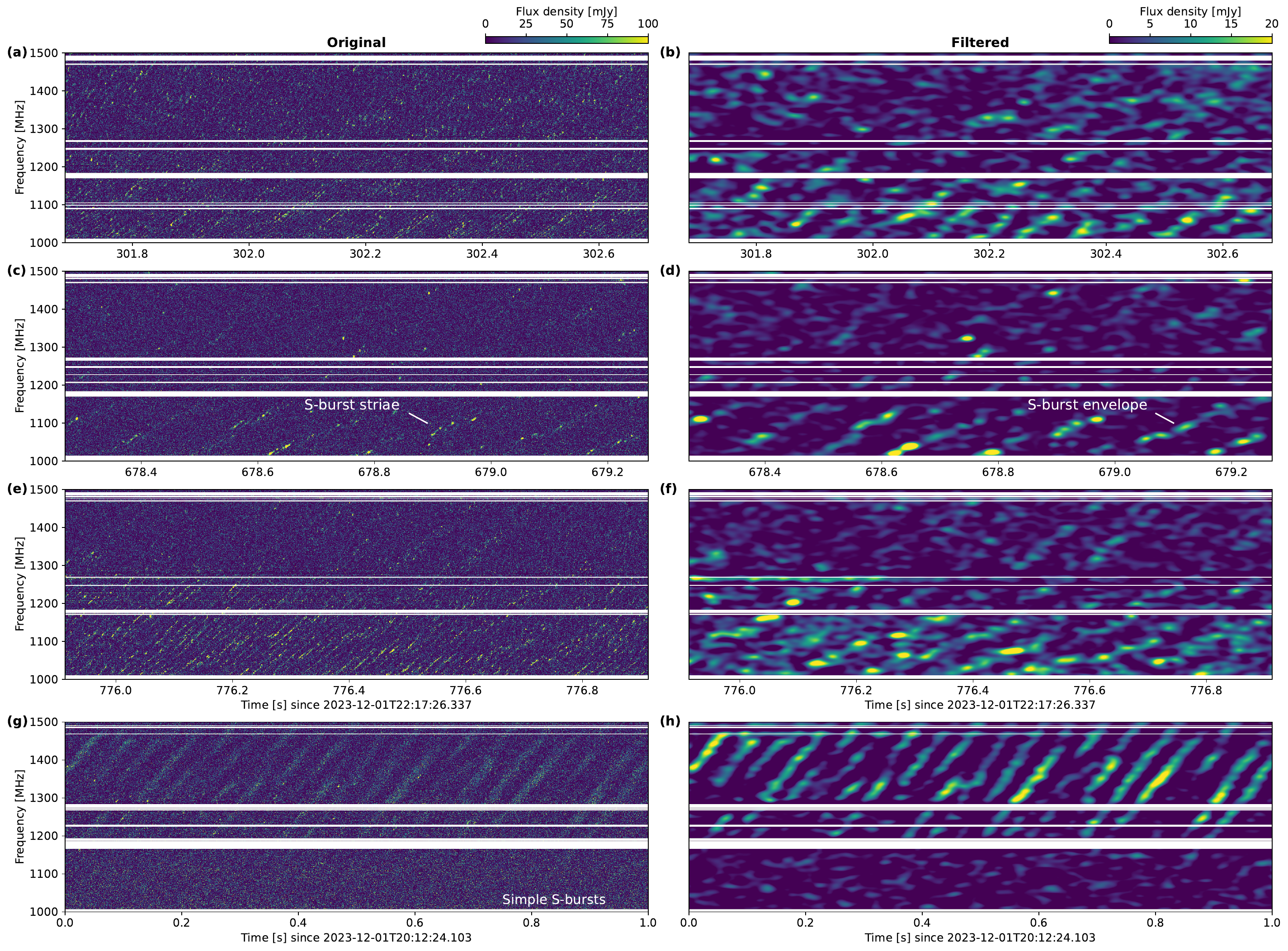}
    \caption{(a, c, e) Examples of S-burst envelopes and striae in the dynamic spectra. The time and frequency resolutions are the same with Figure \ref{fig:figure2}. (b, d, f) Corresponding filtered dynamic spectra after applying the same low pass filter with Figure \ref{fig:figure2}, which reveal the S-burst envelopes. Panel (g, h) is an example of a series of simple S-bursts, occurring around 2 hours before the events in focus.}
    \label{fig:figure3}
\end{figure*}

{Drifting bursts analogous to Jovian S-bursts are a common type of fine structures found in the radio emissions from AD Leo \citep{2023ApJ...953...65Z}. As a representative example of AD Leo S-bursts, we show in Figure \ref{fig:figure3}(g) data taken about 2 hours prior to the main events discussed in this work. We refer to these as simple S-bursts to distinguish them from the more complex forms that will be introduced later. These simple S-bursts exhibit a typical instantaneous bandwidth of $\approx 50$ MHz and a duration of a fraction of a second. They have a uniform frequency drift and a quasi-periodic recurrence of about 0.05$\;$s.} 

{In Figure \ref{fig:figure2}(a, c, e), some frequency-drifting fine structures within the modulation lanes can be discerned. To highlight their prevalence and shared characteristics, we present zoomed-in examples at three different time frames (marked by red arrows in Figure \ref{fig:figure1}(a)) in Figure \ref{fig:figure3}(a, c, e). These emission patterns are more complicated than the simple morphology of S-bursts. The smallest structures resolved appear as narrow striae, with a lifetime of a few milliseconds to tens of milliseconds and an instantaneous bandwidth of about 10 MHz. We refer to these as S-burst striae, as they are smaller compared to the simple S-bursts. These striae also share a consistent frequency sweep and a regular spacing to each other ($\approx$15 ms).}

{In addition, we noticed distinctive clustering of these striae, which emerges as emission envelopes in the filtered dynamic spectra (Figure \ref{fig:figure3}(b, d, f). These elongated envelopes are not artifacts introduced by the low-pass filter. They have a longer duration (up to 0.5 s) and a broader instantaneous bandwidth (about 50\,MHz) compared to the striae. They tend to show up at a period of around 0.1 s. These envelopes display some striking similarities with the simple S-bursts in Figure \ref{fig:figure3}(g), in both size and recurrence, and we call them the S-burst envelopes. The S-burst envelopes have a slightly slower drift rate than the striae, which will be quantified in the next section.}

\begin{figure*}[htbp]
    \centering
    \includegraphics[width=0.65\linewidth]{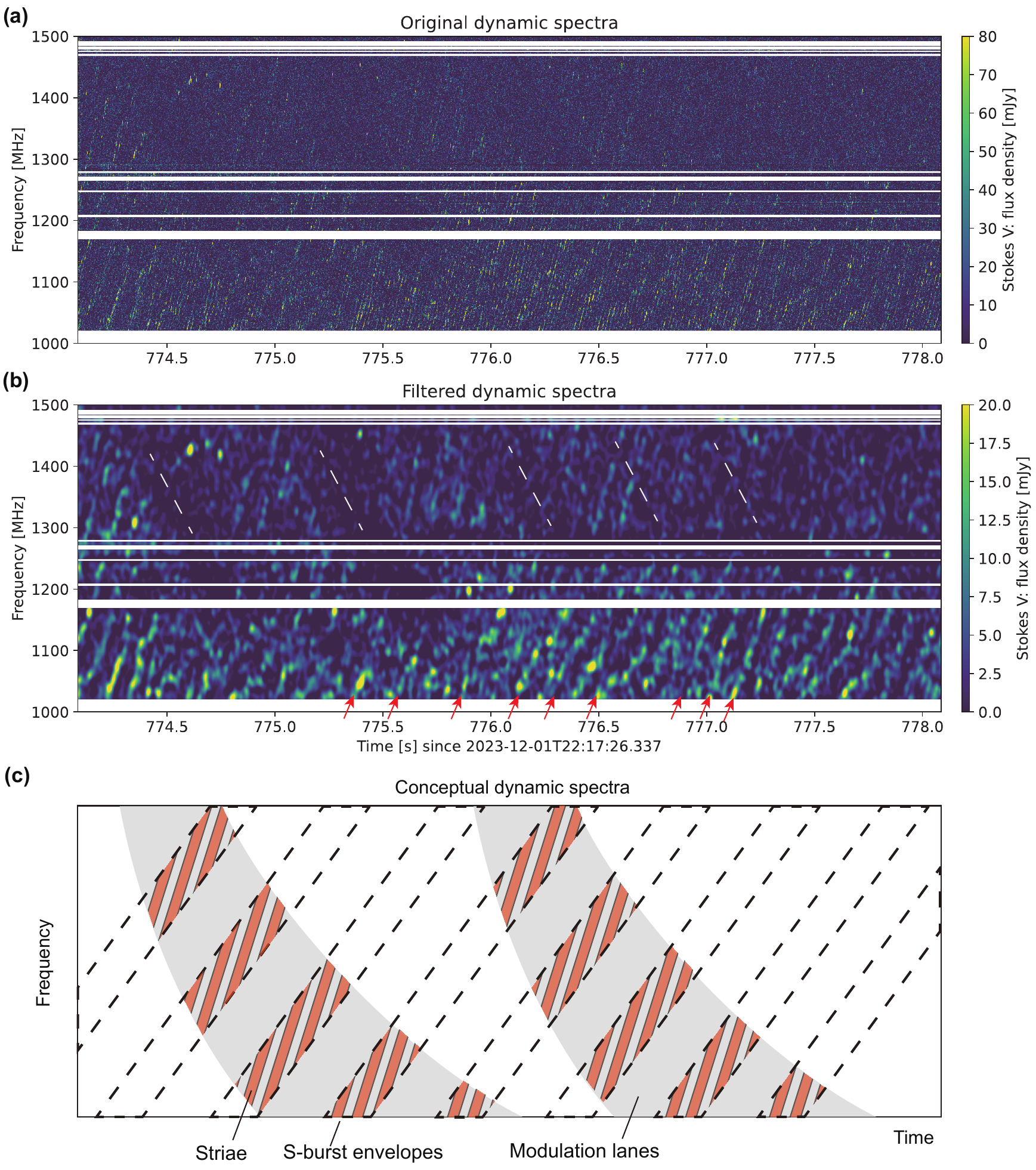}
    \caption{An example of the superposition of modulation lanes, S-bursts and striae in the dynamic spectra. Panels (a) and (b) are the original and low-pass filtered dynamic spectra. The white dashed lines delineate the modulation lanes and the red arrows mark the locations of S-burst envelopes that cross two adjacent modulation lanes. Panel (c) illustrates the topological relationship between the modulation lanes, S-burst envelopes and striae.}
    \label{fig:figure4}
\end{figure*}

{Complex and diverse emission morphologies of Jovian S-bursts have also been reported previously \citep{1999JGR...10425127C,2002JGRA..107.1061W,2009A&A...493..651L,2009GeoRL..3614101H,2014A&A...568A..53R}, although their exact morphological definition  and physical origins remain unclear. So far, we have introduced the three types of emission structures in this burst event: the modulation lanes on second timescales, S-burst envelope and S-burst striae on sub-second to millisecond timescales. Figure \ref{fig:figure4} is a manifestation of the complex superposition of these three different structures. The modulation lanes appear as diffuse, second-long broadband structures with a downward frequency sweep. Two upward-drifting fine structures are the S-burst envelopes and S-burst striae. The S-burst envelopes are better seen in the filtered dynamic spectra (Figure \ref{fig:figure4}(b)) and some of these envelopes are found to cross two adjacent modulation lanes (marked by red arrows). Their topological relationship is illustrated in the conceptual diagram in Figure \ref{fig:figure4}(c).}

The complex superposition of these patterns makes it hard to isolate each spectro-temporal components in the dynamic spectra and {quantify their properties. To further study their internal relationship,} we employed Fourier analysis on the dynamic spectra, using discrete Fourier transform (DFT) and auto-correlation functions (ACFs).

\subsection{Secondary spectra and auto-correlation functions}\label{sec:secondary}
Repeating patterns in the dynamic spectrum can be better identified in the Fourier plane as clustering of power at specific frequency scales. The observed dynamic spectrum is a function of time and frequency, $I(t,\nu)$. The squared modulus of its Fourier conjugate $S(f_t,f_\nu)$, often called the secondary spectra in studies of pulsar scintillations \citep{2001ApJ...549L..97S,2004MNRAS.354...43W,2006ApJ...637..346C}, is expressed as
\begin{equation}
    S(f_t,f_\nu)=\left|\Tilde{I}(f_t,f_\nu)\right|^2=\left|\mathrm{DFT}\left[I(t,\nu)\right]\right|^2,
\end{equation}

\noindent where $f_t$ and $f_\nu$ are the Fourier conjugates of time and frequency, respectively. The two-dimensional (2D) ACF, $A(\Delta t,\Delta \nu)$ is defined by
\begin{equation}
    A(\Delta t,\Delta \nu)=\langle \Delta I(t,\nu)\cdot\Delta I(t+\Delta t,\nu +\Delta \nu)\rangle,
\end{equation}

\noindent where $\Delta I$ is the deviation of the intensity from the mean value, $\langle\rangle$ denotes the average of the product of the deviated intensities over the sampled data, and $\Delta t$ and $\Delta \nu$ are the time delay and frequency delay, respectively. It is known that the secondary spectrum and the ACF form a Fourier pair, which can be written as
\begin{equation}
    A(\Delta t,\Delta \nu)\propto\mathrm{DFT}^{-1}[S(f_t,f_\nu)],
\end{equation}

\noindent where $\mathrm{DFT}^{-1}$ means the inverse DFT. 

Consider a frequency-drifting burst in the shape of an elongated ellipse in the dynamic spectrum. If the burst intensity approximately follows a 2D Gaussian function, its power in the secondary spectrum will also follow a 2D Gaussian function, with the major axis oriented along a perpendicular direction in the Fourier domain. In the case of a group of elliptical bursts occurring at regular time-intervals,  enhanced power will appear in the secondary spectrum at the inverse of the time interval and its harmonics (see Appendix \ref{sec:topology} for more details). The concentration of power at a certain 2D coordinate in the secondary spectra (called hereafter a fringe mode) corresponds to a periodic fringe pattern in the 2D ACF. We present the examples of DFT and ACF diagrams based on 8\;s long dynamic spectra in Figure \ref{fig:figure5}. Flagged channels are treated as zero-valued pixels in the analysis. Due to the conjugate symmetry property of the Fourier transform of real-valued signals, the power in the secondary spectra and ACF is centrally symmetric. 

{We show the secondary spectra of the unfiltered dynamic spectra consisting of S-burst envelopes and striae (Figure \ref{fig:figure5}(a, d, g)), as well as the one including just simple S-bursts (Figure \ref{fig:figure5}(j)) for comparison.} They exhibit highly organized structures, for which we introduce the following morphological terms to facilitate clearer descriptions:

\begin{figure*}[htbp]
    \centering
    \includegraphics[width=0.9\linewidth]{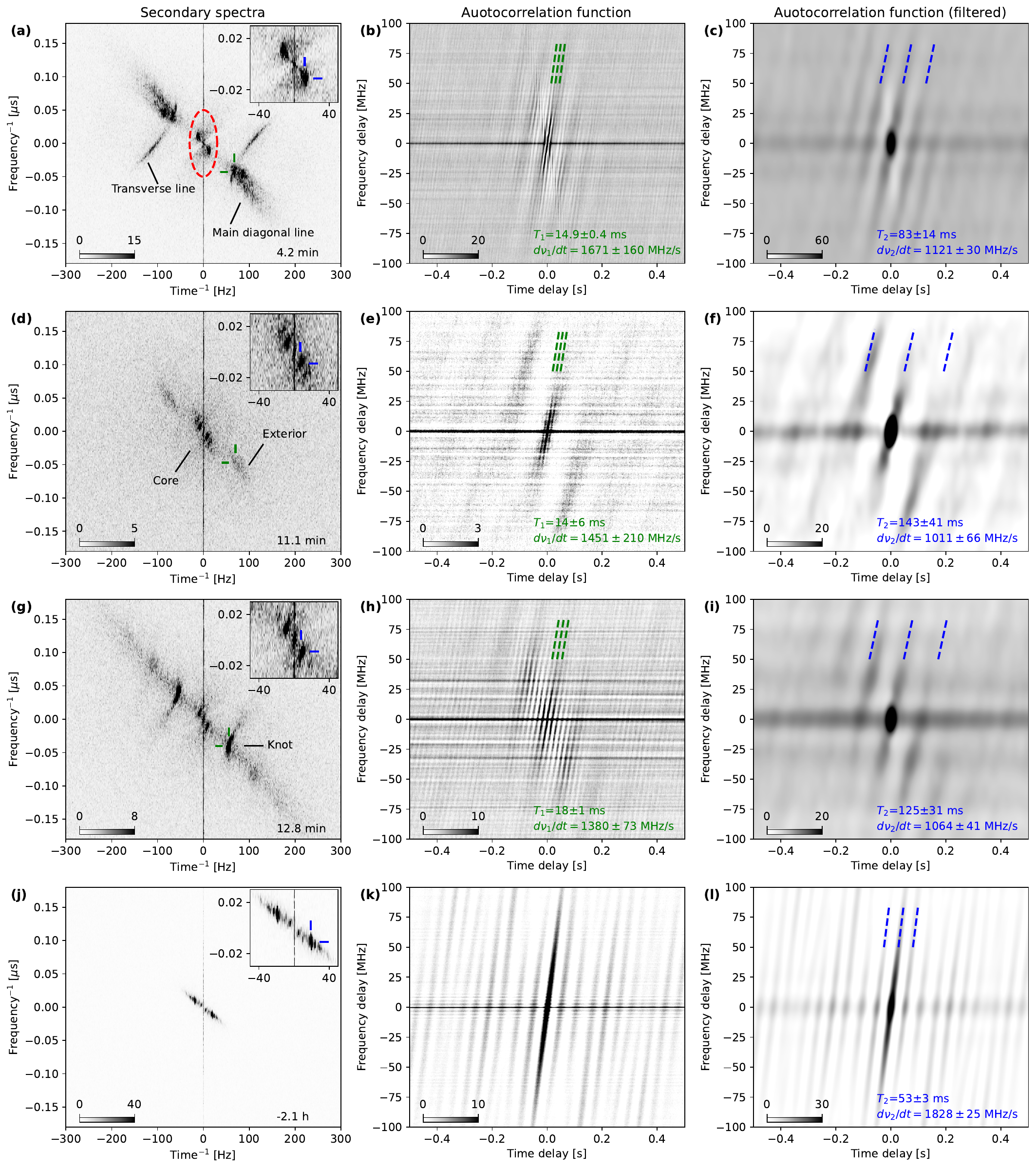}
    \caption{(a, d, g, j) Examples of the secondary spectra based on 8\;s long data. The first 1 s long dynamic spectra of the data have been shown in Figure \ref{fig:figure3}(a, c, e, g). The grey scale is the signal-to-noise ratio (SNR) of power in linear scale. In panel (a), the red dashed ellipse marks the size of the window of the low-pass filter. The sub-panels are the zoom-in core regions, ranging from -50 Hz to 50 Hz and from -0.03 $\mu$s to 0.03 $\mu$s. The blue (green) cross-hairs indicate the strongest low frequency (high frequency) knots in the Fourier plane. (b, e, h, k) ACF of the corresponding original dynamic spectra. The green dashed lines indicate the high frequency fringe mode, with the period and the drift rate marked in the bottom right corner. {These properties are summarized in Table \ref{tab:table1}}. (c, f, i, l) ACF of the filtered dynamic spectra. The blue dashed lines indicate the low frequency fringe mode.}
    \label{fig:figure5}
\end{figure*}

(1) \textit{Core}. The core region represents the region of clustering power at low frequencies in the Fourier plane, defined here as $|f_t|<50$ Hz and $|f_\nu|<0.03$ $\mu$s. The core regions are zoomed-in in the sub-panels of Figure \ref{fig:figure5}(a, d, g, j). The core has an envelope of an oblique ellipse and blobby structures inside.

(2) \textit{Exterior}. The exterior region represents the region outside the core. Certain power patterns in the exterior region are found in Figure \ref{fig:figure5}(a, d, g), while no clear structures can be identified in Figure \ref{fig:figure5}(j). The separation between the core and the exterior power patterns is clearly seen in Figure \ref{fig:figure5}(a, d).

(3) \textit{Knots}. The knots represent the clustering of power at certain locations in the Fourier plane. They are classified into two types, the core knots related to low frequency fringe modes, and exterior knots related to high frequency fringe modes. The exterior knots and the core knots are directly related to the fringe patterns seen in the ACFs of the original dynamic spectra (marked by green dashed lines in Figure \ref{fig:figure5}(b, e, h)) and those in the ACFs of the filtered dynamic spectra (marked by blue dashed lines in Figure \ref{fig:figure5}(c, f, i, l)). The knots are roughly distributed along a diagonal line.

(4) \textit{The main diagonal and the transverse lines}. The main diagonal line refers to the power that lies along the most prominent slope line. The main diagonal line extends from the core region to the exterior region and is seen in all the presented secondary spectra. In some cases, the diagonal line may look ``kinked''. For instance in Figure \ref{fig:figure5} (d), the exterior parts of the main diagonal line do not cross the center. The transverse line refers to the dispersion of power that extends from the knots along a nearly orthogonal direction. The transverse lines are seen in Figure \ref{fig:figure5} (a) and (g), and in Figure \ref{fig:figure5} (a), the transverse line is clearly curved.

(5) \textit{The vertical line}. There is a vertical line crossing the center in all the secondary spectra. This is related to the horizontal line patterns seen in ACFs. It is an artifact from the zero-valued pixels at the flagged frequency channels and will not be discussed further.

The four presented secondary spectra (Figure \ref{fig:figure5} (a, d, g, j) have similar patterns in the core region. The core region represents the low-frequency information, namely structures with bandwidth larger than 20 MHz and a time scale longer than 0.03 s. The power is primarily contributed by variability of the S-bursts or S-burst envelopes. Periodicity of the S-bursts is suggested by the clustering of power at $f_t\approx 10$ Hz, $f_\nu\approx -0.01$ $\mu$s. In the exterior region, however, the first three secondary spectra (Figure \ref{fig:figure5} (a, d, g) reveal very distinctive patterns. Another clustering of power is seen, typically located around $f_t\approx 60$ Hz, $f_\nu\approx -0.04$ $\mu$s. The corresponding time and frequency scales of the structures in the dynamic spectra could only be associated with the S-burst striae, the finest structures that are resolved. The fringe modes of the S-burst envelopes and the striae are best represented in the original ACF and filtered ACF, respectively. Based on the ACFs, we found that the drift rates of the striae ($\approx 1.5$ GHz/s) are relatively faster than the S-burst envelopes ($\approx 1.1$ GHz/s). Spectro-temporal properties of the three spectral structures-- modulation lanes, S-burst envelopes, and S-burst striae are summarized in Table \ref{tab:table1}. The methods used to measure these properties are described in Appendix \ref{sec:drift}.

\begin{figure*}[htbp]
    \centering
    \includegraphics[width=0.9\linewidth]{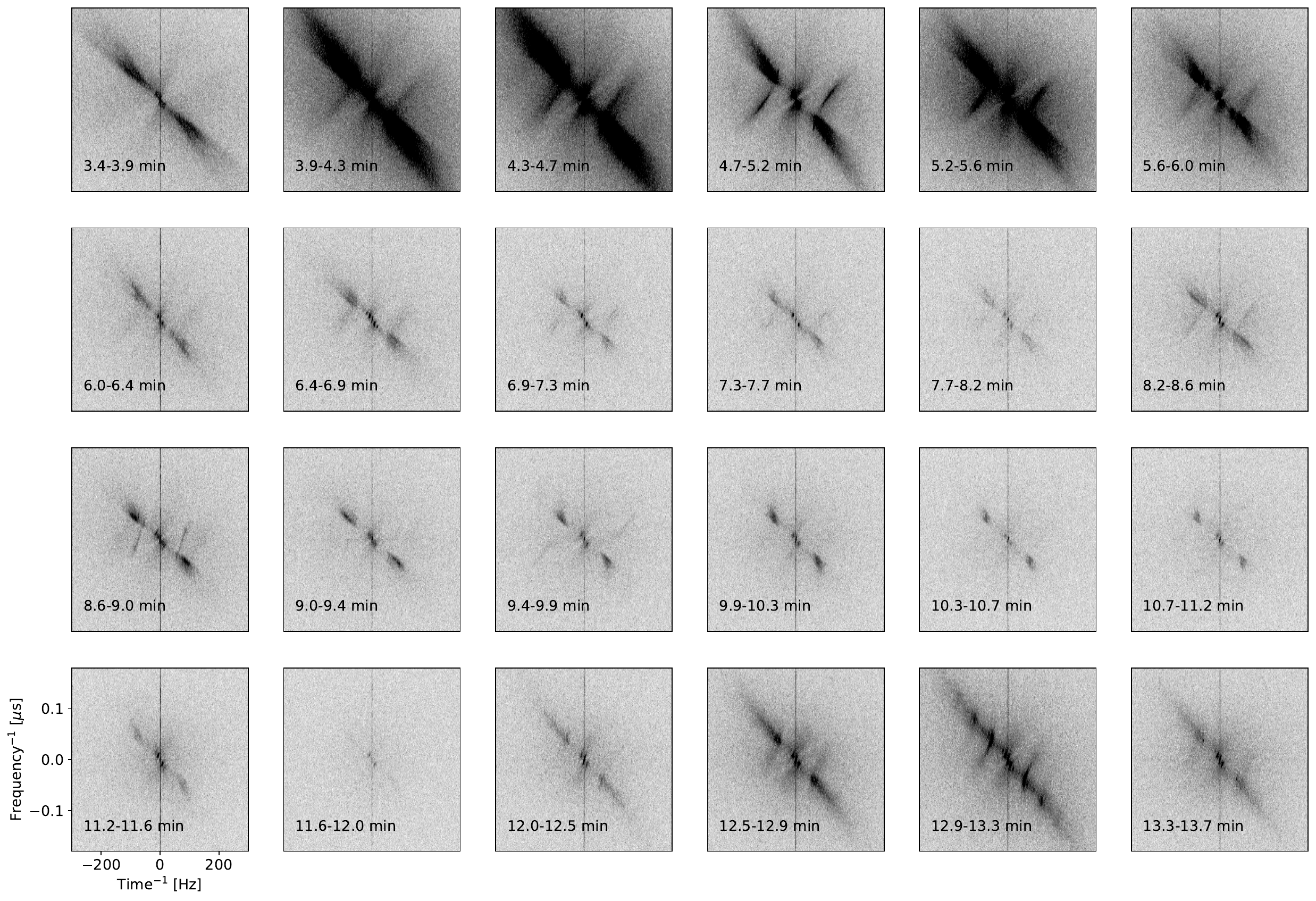}
    \caption{Evolution of the patterns in the secondary spectra. Each secondary spectrum is analyzed based on $\approx 0.4\,$min of data. {The grey scale indicates the SNR of power from 1 to 10 in logarithmic scale, which also represents the relative intensity of the bursts}.}
    \label{fig:figure6}
\end{figure*}

\begin{table*}[htbp]
\centering
\caption{Properties of the modulation lanes, S-bursts, and striae of the events}
\begin{tabular}{c|c|c|c}
& Modulation lanes & S-burst envelopes & S-burst striae \\ \hline
Drift rate           &       $\approx-$ 0.8 GHz/s          &   $\approx$ 1.1 GHz/s       &   $\approx$ 1.5 GHz/s         \\ 
Period                   &  $\approx 1$ s or random  &   80 - 150 ms       &   14 - 18 ms         \\ 
Instantaneous bandwidth   &      $>$ 400 MHz            &    $\approx$ 50 MHz     &     $\approx$ 10 MHz      \\ 
Single-frequency duration &      0.5 - 2 s            &     40 - 75 ms     &   7 - 9 ms         \\ \hline
\end{tabular}
\label{tab:table1}
\end{table*}

A 10-min evolution of the patterns in the secondary spectra is shown in Figure \ref{fig:figure6}. {Though the patterns change significantly over the 10 mins, all the structures presented here could be described using the terminology that we have introduced before. We are particularly interested in the structures in the exterior region, which we find most challenging to interpret from drifting bursts}. We developed a mathematical model in Appendix \ref{sec:topology} to simulate different types of patterns in the dynamic spectra which are simply composed of elementary bursts with a uniform shape. We show five types of dynamic spectra and their corresponding secondary spectra, which we considered as trivial, including random blob bursts, random stripe bursts, periodic stripe bursts, random curved bursts, periodic curved bursts. The observed secondary spectra display some unique structures that could not be topologically accommodated in this simple framework. The closest approximation we achieved was by multiplying the original burst pattern with another periodic modulation pattern, which seems to reproduce the structures observed in the secondary spectrum of Figure \ref{fig:figure5}(d) (see Appendix \ref{sec:topology}). This implies that the observed dynamic spectra cannot be regarded as a collection of elementary bursts with quasi-uniform shape. A convolution of at least two distinct time-frequency structures has to be introduced to account for all the characteristics.

\section{Interpretations of patterns in the radio emission of AD Leo}\label{sec:interpretations}

During the 15-min observation, AD Leo displays emission variability across timescales from seconds down to milliseconds. There are three distinct types of emission patterns visible in the dynamic spectra, two of which, the modulation lanes and the S-burst striae, have not been previously investigated. The resemblance between the S-burst envelopes and the simple S-bursts suggests a common intrinsic origin, which fits in a classical framework for drifting bursts \citep{1996GeoRL..23..125Z}. However, the physical origin of the two additional structures with distinct drift rates remains unclear. Furthermore, the current understanding of drifting S-bursts fails to account for the unusual features observed in the secondary spectra. These limitations prompt us to consider the potential role of propagation effects in adding another layer of structures in the observed emission. In the following sections, we explore possible interpretations for each of the three structures. Briefly, we interpret the S-burst envelopes as being intrinsic to the source and the modulation lanes as magnification of intensity due to refraction from an inhomogeneous plasma medium along the line-of-sight (LOS). S-burst striae are the most puzzling structures: although we attempt to model them as a result of diffractive scatterings, we do not find a satisfying explanation for their time-frequency scales.

\subsection{S-burst envelopes}

The S-burst envelopes show similarities with millisecond radio bursts that have been reported on AD Leo, suggesting radio emission produced by ECM mechanism \citep{2008ApJ...674.1078O,2023ApJ...953...65Z}. For simplicity of the following discussion, fundamental emission mode is assumed. We believe that each S-burst envelope corresponds to an electron charge bunch that is moving along a set of magnetic field lines. Hereafter, we call the radio-emitting field lines as active field lines (AFLs). As a result, the drift rate of the S-burst is determined by the parallel velocity of the electrons, $v_e$ and the magnetic field gradient, $\nabla_\parallel B$ along the AFLs,

\begin{equation}
    \left(\frac{d \nu}{d t}\right)_{\rm burst}=2.8 \mathrm{\;MHz/G\;}\cdot v_e \nabla_\parallel B.
\end{equation}

As the drift rate is positive, the electrons are moving towards lower altitudes with stronger magnetic field strengths. We define a magnetic scale length along the AFLs as $L_B=B/\nabla_\parallel B$. Substituting $\nabla_\parallel B$, the normalised drift rate is given by
\begin{equation}
    \frac{1}{\nu}\left|\frac{d \nu}{d t}\right|_{\rm burst}=\frac{v_e}{L_B}.
\end{equation}

{ECM emission is usually associated with weakly or mildly relativistic electrons \citep{2006A&ARv..13..229T}, while the more energetic electrons are responsible for gyrosynchrotron or synchrotron radiations.} Considering an S-burst with a drift rate of $1.1$ GHz/s and assuming a typical electron energy of 2 - 100 keV (or 0.1 - 0.6 c in velocity based on planetary observations, \cite{1998JGR...10320159Z,2023JGRA..12831985L,2023NatCo..14.5981M}) responsible for the ECM emission, the magnetic scale length could be constrained to $L_B= 0.1 -0.5\; R_*$ ($R_*$ as the stellar radius). The periodicity of the S-bursts might be related to the periodic modulation on the velocity distribution of the energetic electrons from a certain emission driver (e.g. kinetic Alfvén waves, \cite{2007JGRA..11211212H,2007ApJ...665L.171W}).

It is worth noting that, for a broadband ECM emission, the emissions of different frequencies are coming from different locations of AFLs. This frequency–to-position mapping is an important consideration and will be revisited in the following discussion of propagation effects.

\subsection{Modulation lanes}

Propagation effects like scintillation could exert an intensity modulation independent of the intrinsic burst patterns. It is possible that there are coherent time-frequency structures that could arise from the radio wave scattering in the corona. We first examine if they may share similar characteristics with the observed modulation lanes. We assume that the scattering medium is composed of plasma material trapped by the magnetic field of the star. Under this assumption, the plasma screen should be located at the sub-Alfvénic regime of the star, possibly within a few or tens of stellar radii according to some wind modellings of magnetically-active M dwarfs (e.g. \cite{2021MNRAS.504.1511K,2022ApJ...928..147A,2024ApJ...971..153X}). Furthermore, we assume that the thin screen approximation is applied for the propagation effect, where the scattering of radio waves predominately occurs at a very thin layer of plasma and the rest of the propagation path is considered to be non-scattering. 

Let $z$ be the distance between the plasma screen and the source. $\boldsymbol{x}$ denotes the position on the screen plane and $\boldsymbol{r}$ denotes the position on the source plane. Both planes are defined to be orthogonal to the LOS. The electric field received by the observer is described by the Fresnel-Kirchhoff integral (F-K integral) under the near-field diffraction condition

\begin{equation}
    u(\boldsymbol{r},\nu) =\frac{-i}{2\pi r_F^2}\iint_{\rm screen} d^2\boldsymbol{x} \; \exp{\left[i\Phi(\boldsymbol{x},\boldsymbol{r},\nu)\right]},
\end{equation}

\noindent where the total phase $\Phi(\boldsymbol{x},\boldsymbol{r},\nu)$ is given by
\begin{equation}
    \Phi(\boldsymbol{x},\boldsymbol{r},\nu)=\frac{|\boldsymbol{x}-\boldsymbol{r}|^2}{2r_F^2}+\phi(\boldsymbol{x},\nu),
\end{equation}

\noindent $r_F$ is the Fresnel scale and $\phi(\boldsymbol{x},\nu)$ is the phase variation related to the spatially-varying refractive index of the plasma screen. We define the magnification/de-magnification of the apparent intensity of the emission at a given source location $\boldsymbol{r}$ and frequency $\nu$ as the modulation pattern, determined by $|u(\boldsymbol{r},\nu)|^2$. When the emission frequency is much higher than the ambient plasma frequency $\nu\gg\nu_{pe}\approx 8.98\times 10^3\cdot \sqrt{n_e}\;\mathrm{[Hz]}$ (requiring plasma density $n_e\ll10^{10}\mathrm{cm}^{-3}$ at our observing frequency), the phase variation is given by

\begin{equation}
    \phi(\boldsymbol{x},\nu)=-r_e\lambda \Delta N_e(\boldsymbol{x}).
\end{equation}

Here, $r_e$ is the classical electron radius $r_e=e^2/(m_e c^2)\approx 2.82\times 10^{-13}\;$cm ($e$ as the electron charge and $m_e$ as the electron mass), $\lambda$ is the observed wavelength, $\Delta N_e$ is the variation of the electron column density: $\Delta N_e=\int_{\rm screen} \Delta n_e dl$. 

In the strong scattering regime, where the phase variation of the plasma screen is much greater than 1\;rad across the Fresnel scale, multiple scattered images emerge on the plasma screen. These images are also called the stationary phase points (SPPs) \citep{2006ApJ...647.1131M} and represent locations where the first-order derivative of the total phase is zero,
\begin{equation}
    \nabla_{\boldsymbol{x}}\Phi(\boldsymbol{x},\boldsymbol{r},\nu)=0.
\end{equation}

The F-K integral can be approximated by summing contributions from the neighbourhoods of SPPs. Two general types of intensity modulation can arise: one due to interference between multiple images, known as diffractive scattering; the other due to changes in the spatial distribution of these images, referred to as refractive scattering. If two images coalesce at a point where the second-order derivative of the phase approaches zero, $\nabla_{\boldsymbol{x}}^2\Phi=0$, a (fold) caustic forms \citep{2006ApJ...647.1131M,2006ApJ...647.1142W}, resulting in strong magnification of the source.

\subsubsection{A sinusoidal plasma screen}
We explore the possibility that the observed modulation lanes may arise from caustics. To simulate the modulation pattern and pseudo dynamic spectrum, a specific realisation of the plasma screen is required. As a starting point, we consider the simplest case: a plasma screen with uniform density along one axis (the $y$-axis) and a sinusoidal variation along the other axis (the $x$-axis). This one-dimensional (1D) sinusoidal plasma screen is adopted as a heuristic model to understand emission patterns and spectral characteristics due to scintillation. The resulting phase variation introduced by the screen can be expressed as:

\begin{equation}
    \phi(x,\nu)=\phi_0(\nu)\cos(q_0x),
\end{equation}

\noindent where $\phi_0$ and $q_0$ are the amplitude and spatial wavenumber of the phase variation. $\phi_0$ is determined by the emission wavelength and fluctuation of the column density in the screen $\Delta N_0$, $|\phi_0|=r_e\lambda\Delta N_0$. We define the density inhomogeneity scale of the screen as $x_0=2\pi/q_0$. The 1D sinusoidal screen is one of the few idealized cases where a closed-form solution of the F-K integral exists \citep{2006ApJ...647.1131M,2006ApJ...647.1142W}.  The analytical solution to the F-K integral of the sinusoidal screen is given by \citep{1997RaSc...32..913B,2006ApJ...647.1131M}
\begin{equation}
    u(r,\nu)=\sum_{n=-\infty}^{\infty}i^n J_n(\phi_0)\exp\left[in(q_0r-nq_0^2r_F^2/2)\right],
\end{equation}

\begin{figure*}[htb!]
    \centering
    \includegraphics[width=0.7\linewidth]{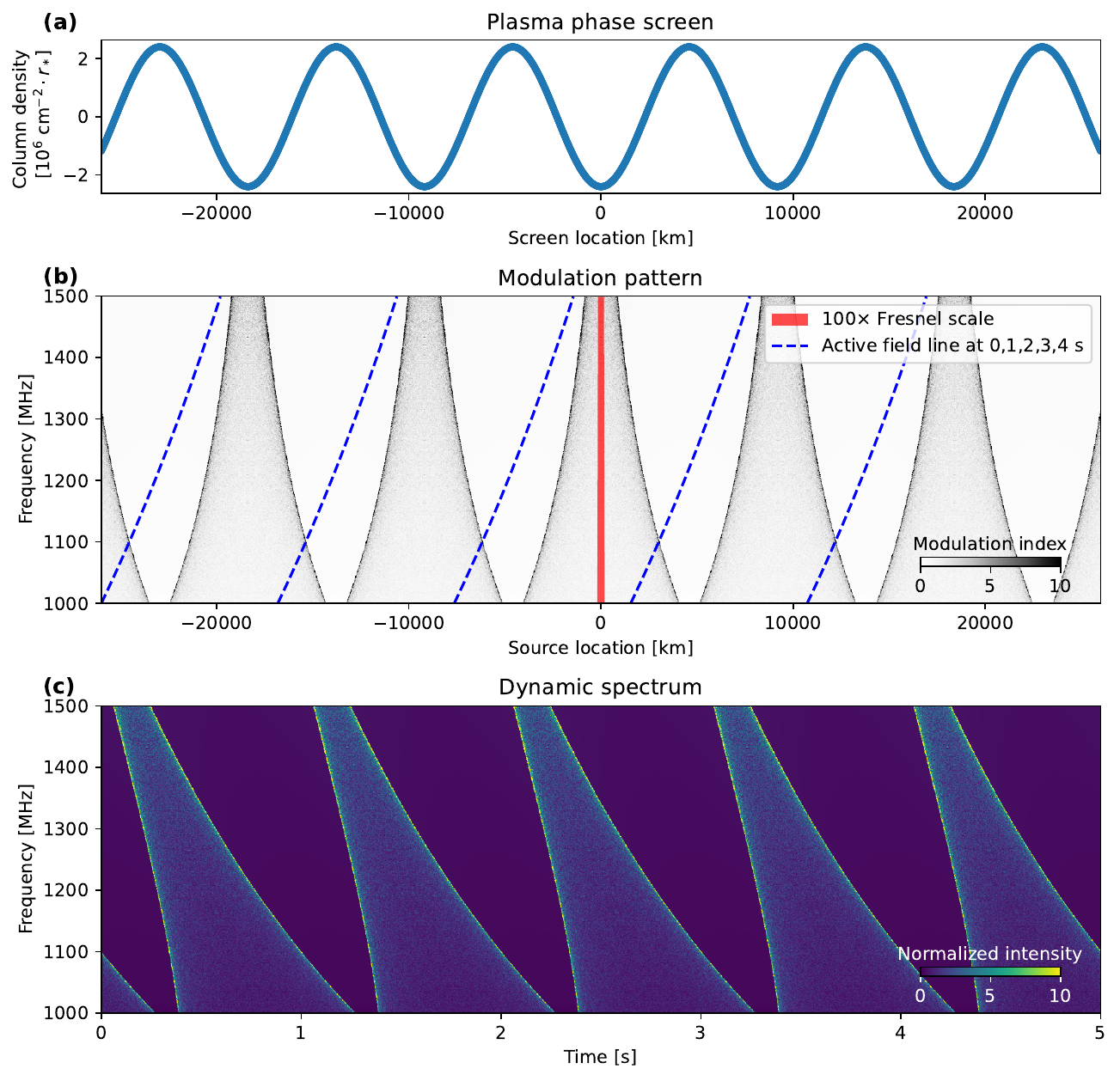}
    \caption{Simulation of the AFLs passing through the modulation pattern of a sinusoidal plasma phase screen. (a) Column density variation of the plasma screen. (b) Modulation pattern. The grey scale represents the modulation index—values greater than one indicate magnification. The red region corresponds to 100 times the Fresnel scale, and the blue dashed lines indicate the AFL positions at different time frames (0, 1, 2, 3, 4 s from left to right). (c) Simulated dynamic spectrum. The colour scale corresponds to the intensity normalized to a mean value of 1.}
    \label{fig:figure7}
\end{figure*}

\noindent$J_n$ is the Bessel function of order $n$. The properties of the modulation patterns from scattering of a 1D sinusoidal screen have been thoroughly discussed by \cite{2006ApJ...647.1131M} and \cite{2006ApJ...647.1142W}, therefore we do not elaborate further. In Figure \ref{fig:figure7}(a,b), a specific example is presented, with a plasma screen at a distance of $z=R_*$, {an inhomogeneity scale of $x_0=0.03\, R_*$, and a column density fluctuation of $\Delta N_0=2.4\times 10^6\,$ cm$^{-3}\cdot R_*$. This corresponds to an amplitude of differential phase of $|\phi_0|\sim5\times 10^5\,$ rad and a root-mean-square phase variation across the Fresnel scale of $\zeta\sim\phi_0r_F/x_0\sim200\;\rm rad$ at our observing frequencies.} In this case, a maximum of three images are created. Caustics correspond to the source locations shown as curved black lines in Figure \ref{fig:figure7}(b), where the magnification reaches the maximum. In between two adjacent caustics are grey regions where the emission is moderately amplified due to focusing of radio waves from a local low-density region (as the plasma lens acts as a diverging lens). Meanwhile, sources projected at high-density region will have reduced apparent intensity due to defocusing (white regions in Figure \ref{fig:figure7}(b)). Besides, the interference between the images results in finer modulation fringes due to diffractive scattering (not resolved in Figure \ref{fig:figure7}(b)), which will be discussed in the next section with regard to the striae.

\begin{figure*}[htb!]
    \centering
    \includegraphics[width=0.95\linewidth]{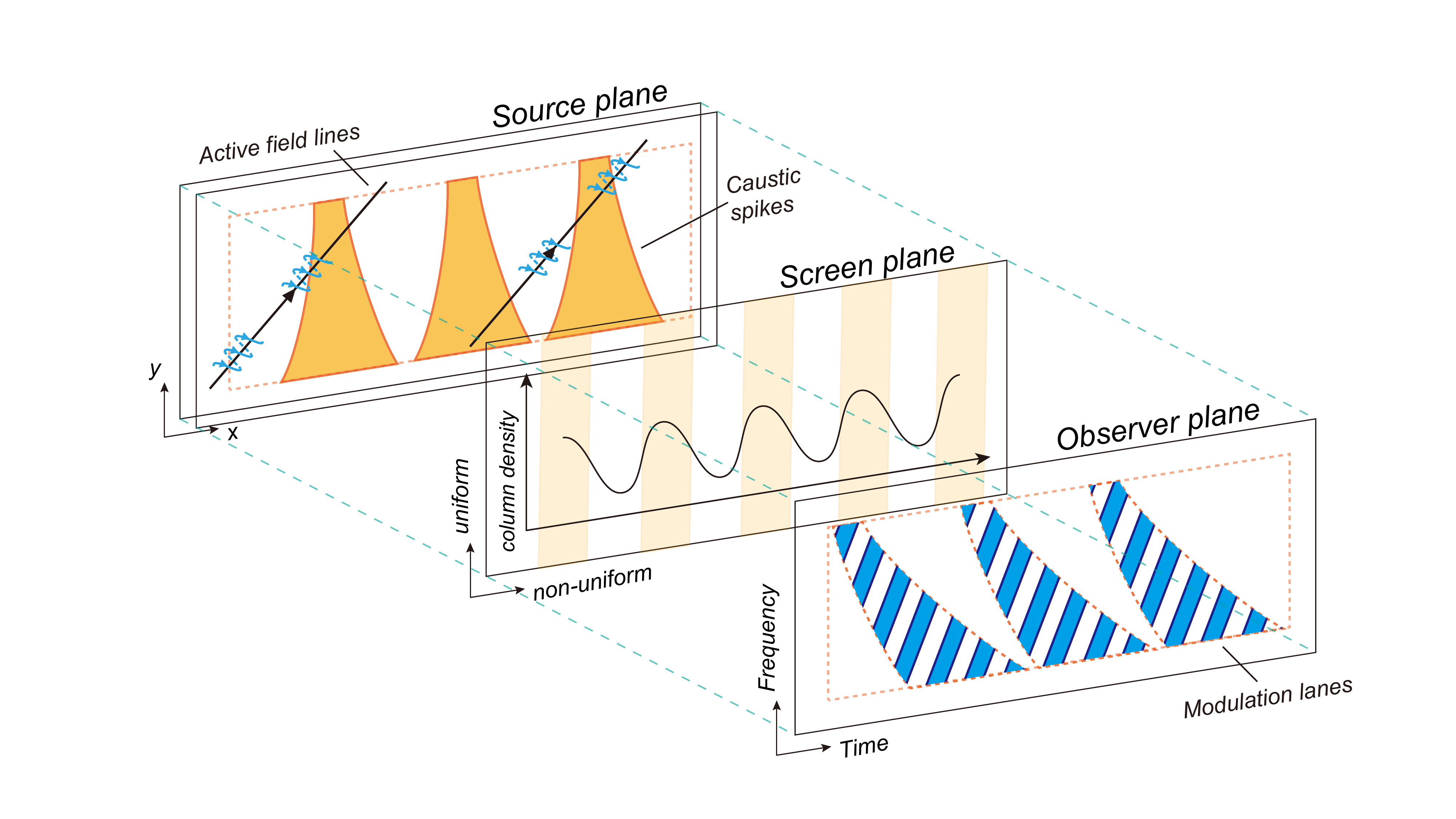}
    \caption{Schematic diagram of the formation of the modulation lanes from the refraction of a sinusoidal plasma phase screen. On the source plane, two black lines show AFLs moving from left to right with respect to the screen. The blue helixes represent the elementary emitters (electron charge bunches) producing the S-bursts. The triangular orange regions indicate the magnification regions in the modulation pattern. On the screen plane, the orange vertical columns show regions that are relatively overdense. The black curve denote the column density variation. On the observer plane, a conceptual dynamic spectrum is illustrated.}
    \label{fig:figure8}
\end{figure*}

Temporal modulation of the intensity could be interpreted as the relative motion between the source and the plasma screen. We take the plasma screen as the frame of reference and consider radio-emitting AFLs moving across a static modulation pattern. As illustrated in Figure \ref{fig:figure8}, the intensity of the AFLs is modulated as a function of time and frequency as they move in and out of these magnification regions. The time axis in the dynamic spectra could be transformed to a position offset of the sources from the modulation pattern. This physical picture accounts for some key properties of the observed modulation lanes. Firstly, it explains that in many cases, the edges of the modulation lanes are quite distinctive, especially in Figure \ref{fig:figure1}(c) and Figure \ref{fig:figure2}. This is because that the edges correspond to the source crossing caustics, bringing sudden and significant changes to the magnification. Secondly, it explains at least qualitatively that almost all the modulation lanes have a decreasing width towards higher frequencies and {vice} versa for the gaps in between. The temporal width of the modulation lanes is related to the distance between two adjacent caustics, which is frequency-dependent as shown in Figure \ref{fig:figure7}(b). 

The frequency sweep of the central lines of the modulation lanes can be explained by the frequency-to-position mapping of the AFLs. This is a unique effect not present in canonical scintillation of instantaneously-broadband radio sources such as pulsars \citep{2004MNRAS.354...43W}. As the radio emissions at different frequencies originate from different parts of the magnetic field lines, there is a time delay when they arrive at the same $x$-axis position on the modulation pattern. In Figure \ref{fig:figure8}, sources emitting at different frequencies have different $x$-axis position offsets. The dynamic spectrum is an affine transformation of the modulation pattern. As there are also compact sources moving along the AFLs and producing narrowband and discrete bursts, S-bursts are seen as an additional pattern to the modulation lanes. If the compact source travels across two magnification regions, it will appear in two adjacent modulation lanes (as demonstrated in Figure \ref{fig:figure4}).

The spectro-temporal structures of the modulation lanes impose joint constraints on the bulk velocity, inhomogeneity scale, distance, and density fluctuation of the plasma screen. {The typical temporal separation between two lanes (e.g. leading edge to leading edge)} $\Delta t$ is around 1 s in Figure \ref{fig:figure2}(c). Suppose that the AFLs are travelling at a transverse velocity (only the $x$-axis component is important) of $v_t$ relative to the scintillation pattern, 
\begin{equation}
    v_t \approx {x_0}/{\Delta t}.
\end{equation}

The modulation lanes are observed to drift at a rate of $\approx 0.8\,$ GHz/s, which corresponds to
\begin{equation}
    \frac{1}{\nu}\left(\frac{d\nu}{dt}\right)_{\rm lanes}\approx\frac{v_t}{L_B\sin\alpha},
\end{equation}

\noindent where $\alpha$ is the angle between the AFLs and the $y$-axis, along which the column density is assumed to be uniform. If the AFLs are perfectly aligned with the $y$-axis, there will be no frequency-related offset and the central line of the modulation lanes will appear vertical in the time-frequency plane ($(d\nu/dt)_{\rm lanes}$ reaching infinity). Combining equations (12) and (13) and taking characteristic frequency of $\nu= 1.25\,$ GHz, yields $x_0\approx 0.6\, L_B\sin \alpha$. This means that the spatial scale of the inhomogeneity should be close to one half of the projected magnetic scale length. For a constrained magnetic scale length of $L_B=(0.1-0.5)\;R_*$ and a possible angle of $\alpha=1 ^{\circ}-90^{\circ}$, the inhomogeneity scale falls in the range of $x_0\approx(0.001-0.3)\;R_*$. {For the specific case used in the simulation with $x_0=0.03\,R_*$, a transverse velocity of $v_t\approx9\times 10^3\,$ km/s and a projected magnetic scale length $L_B\sin\alpha\approx1.5\times10^4\,$km for the AFLs is required to account for the time scale and drift rate of the observed modulation lanes.} Such high velocity might correspond to the propagation phase velocity of the density structures of the plasma screen, which will be discussed later. If the inhomogeneity scale is set larger or smaller, the transverse velocity and the projected magnetic scale length need to be scaled accordingly to maintain the same pattern in the dynamic spectra (Figure \ref{fig:figure9}(a)).

To simulate a pseudo dynamic spectrum, we modelled a radio-emitting thin AFL with emission frequency following an exponential relationship with $x$-axis position according to the derived projected magnetic scale length. The resulting dynamic spectra is shown in Figure \ref{fig:figure7}(c). The simulated emission pattern is very similar to the quasi-periodic ($\approx 1\;\rm s$) modulation lanes presented in Figure \ref{fig:figure2}(c,d). In other cases (Figure \ref{fig:figure2}(a,b,e,f)), the modulation lanes are more random and variable, implying that the real density fluctuation could be much more stochastic compared to idealized sinusoidal variation. 

Furthermore, we can constrain the plasma density fluctuation of the plasma screen by considering the single-frequency duration of the modulation lanes, which correspond to the separation of adjacent caustics. We introduce the dimensionless parameter $A=\phi_0q_0^2r_F^2$ similar to \cite{2006ApJ...647.1131M,2006ApJ...647.1142W}. Caustics only occur when $|A|>1$. The distance between the two caustics from the same density trough is given by \citep{2006ApJ...647.1131M,2006ApJ...647.1142W}
\begin{equation}
    \Delta r=\frac{x_0}{\pi}\left[\cos^{-1}\left(\frac{1}{|A|}\right)-|A|\sin\left(\cos^{-1}\left(\frac{1}{|A|}\right)\right)\right],
\end{equation}

\noindent with $\cos^{-1}(1/|A|)$ taking the minimum positive value. We put an upper limit on $|A|$ by assuming $\Delta r\lesssim x_0$, which means that the modulation lanes are just separated from one another. The joint constraint is $1<|A|\lesssim 4.6$. 

To estimate the local plasma density fluctuation, we assume that the LOS extent of the plasma screen is the same with the transverse inhomogenuities scale. The results are shown in Figure \ref{fig:figure9}(b). Qualitatively, a larger inhomogeneity scale or a smaller distance indicates a larger refraction angle, which requires a stronger plasma lens with a larger density gradient. {If the assumption holds, the adopted model in Figure \ref{fig:figure7} requires a local density fluctuation of $\approx8\times10^7\;\mathrm{cm}^{-3}$, which sets the lower limit on the possible local plasma density at the plasma screen region.}

{To compare these estimated values with the typical plasma environment of the AD Leo corona, we modeled the magnetic field and plasma density profiles as functions of radial distance from the stellar center $R$ (Figure \ref{fig:figure9}(c,d)). A dipolar field, $B = B_0 (R / R_*)^{-3}$, was adopted as a first-order approximation to describe the magnetic field variation with height, assuming a surface field strength of $B_0 = 920\;\mathrm{G}$ \citep{2023A&A...676A..56B}. Based on this model, the radio source is located at approximately $1.3\;R_*$, the plasma screen at about $2.3\;R_*$ (assuming the one stellar radius separation) (Figure \ref{fig:figure9}(c)).}

The plasma density distribution was inferred from hydrostatic equilibrium model under constant gravity \citep{2019ApJ...871..214V,2023ApJ...953...65Z}. The density is assumed to decay exponentially at a rate described by density scale height,
\begin{equation}
    n_e=n_{e0}\exp \left[ -(R-R_*)/H_n \right],\quad H_n=\frac{k_BT}{\mu m_H g},
\end{equation}

where $n_{e0}$ is the base coronal plasma density, $k_B$ is the Boltzmann constant, $T$ is the coronal temperature (we adopted three distinct temperatures $T=3\,$ MK, $T=6\,$ MK, $T=9\,$ MK from the temperature range of the peak X-ray emission measure $T=2-10 \;\mathrm{MK}$ \citep{2018ApJ...862...66W}), $\mu$ is the mean molecular weight (we used the solar value 0.6), $m_H$ is the mass of the hydrogen atom, and $g$ is the surface gravity {(log $g=4.8$ [cm/s$^2$], \cite{2015A&A...577A.132M})}. {A typical coronal base density of $10^9\;\mathrm{cm^{-3}}$ is adopted, which is largely motivated by the requirement to generate radio emission that is nearly 100$\%$ circularly polarized at x-mode \citep{2025A&A...695A..95Z} and allow it to escape. This condition corresponds to $\nu_{pe}/\nu_{ce}\lesssim 0.3$ at the source location (illustrated by the dashed lines in Figure \ref{fig:figure9}(d)). Such low densities are consistent with models suggesting that ECM emission originates predominantly within low-density cavities in the stellar magnetosphere \citep{2019MNRAS.488..559Z,2025A&A...695A..95Z}. In contrast, X-ray observations usually infer a much higher coronal base density (up to $10^{10}\;\mathrm{cm^{-3}}$, \cite{2004A&A...427..667N}), possibly implying a highly inhomogeneous coronal environment. The lower limit from the density fluctuation of the plasma screen ($n_e>8\times10^7\;\mathrm{cm^{-3}}$, the red arrow in Figure \ref{fig:figure9}(d)) seems to favor a coronal environment with a higher temperature and a larger density scale height, but it is also possible that the plasma screen corresponds to a significantly overdense region that is an order of magnitude denser than the ambient plasma.}

\subsubsection{Astrophysical implications}

\begin{figure*}[htb!]
    \centering
    \includegraphics[width=0.9\linewidth]{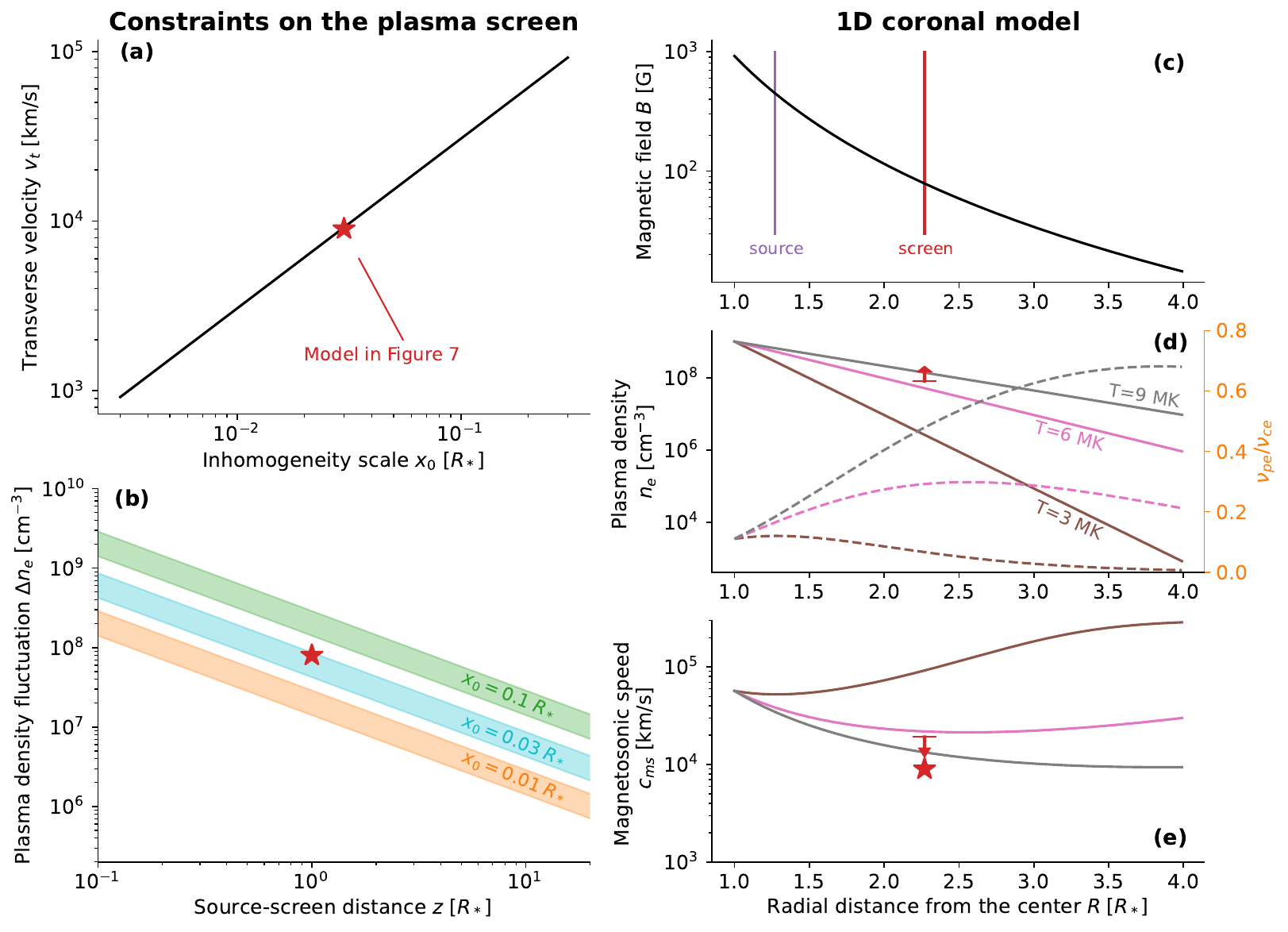}
    \caption{{(a, b) Constraints on the properties of the plasma screen from the observed modulation lanes. Panel (a) shows the relative transverse velocity $v_t$ between the plasma screen and the AFLs as a function of the inhomogeneity scale $x_0$. Panel (b) shows the joint constraints on the plasma density fluctuation $\Delta n_e$, source-screen distance $z$ and inhomogeneity scale of the plasma screen $x_0$. The green, cyan and orange colors represent $x_0=0.1\,R_*$, $0.03\,R_*$, $0.01\,R_*$, respectively. The red stars in the two panels represent the parameters used in the model of Figure \ref{fig:figure7}. (c, d, e) The profiles of the magnetic field strength $B$, plasma density $n_e$, and fast-mode magnetosonic wave velocity $c_{ms}$ as a function of radial distance from the stellar center. In panel (c), the magnetic field strength follows a $r^{-3}$ decay assuming a dipolar field. The locations of the radio source and the plasma screen are indicated. Panel (d) shows the plasma density distributions in solid lines (brown, pink and grey) under three coronal temperatures (3 MK, 6 MK and 9 MK) assuming hydrostatic equilibrium condition. The dashed lines indicate the corresponding ratio between the plasma frequency $\nu_{pe}$ and the cyclotron frequency $\nu_{ce}$. The arrow represents the lower limit on the plasma density at the screen location given by the density fluctuation of the plasma screen. Panel (e) shows the estimated magnetosonic speed under the three coronal temperatures. The red arrow represents the upper limit on the local magnetosonic speed given the lower limit on the local plasma density. The red star is the transverse velocity of the plasma screen from the model in Figure \ref{fig:figure7}.}}
    \label{fig:figure9}
\end{figure*}

The introduced plasma screen is most consistent with magnetically-confined plasma material within the stellar magnetosphere. AD Leo is known to have a dominant dipolar magnetic field, with large-scale surface field strength reaching up to $\sim$900 G at the magnetic pole \citep{2008MNRAS.390..567M,2018MNRAS.479.4836L,2023A&A...676A..56B}. The strong global magnetic field may support localized overdense structures, like plasma trapped by magnetic loops, that extends to a few stellar radii or possibly even further. Density inhomogenuities in such regions might be regulated by magnetohydrodynamic waves. For instance, the phase velocity of a fast-mode magnetosonic wave propagating transverse to the magnetic field is given by $c_{ms}=\sqrt{c_A^2+c_s^2}$, where $c_A$ is the Alfvén velocity ($c_A=v_A/\sqrt{1+v_A^2/c^2}$, $v_A=B/\sqrt{4\pi\rho}$, $B$ and $\rho$ are the local magnetic field strength and plasma density) and $c_s$ is the sound speed ($c_s=\sqrt{\gamma k_BT/(\mu m_H)}$, $\gamma=5/3$ is the adiabatic index). {In the modeled coronal environment, the velocity is dictated by the Alfvén velocity, reaching up to a fraction of the speed of light (Figure \ref{fig:figure9}(e)). The required velocity is a few times smaller than the estimated values from different density profiles, but is generally consistent with the upper limit given by the density constraints of the screen (the red arrow in Figure \ref{fig:figure9}(e)). The discrepancy might be attributed to an over-density of the region, which decreases the local magnetosonic speed.}

The presence of the magnetic field also provides a natural explanation for the high anisotropy of the inhomogeneity of the assumed plasma screen. The density variation is assumed to be one-dimensional, which is most likely to be perpendicular to the magnetic field. This is because that, under the frozen-in plasma condition, plasma could flow freely along the magnetic field lines but not across them,  resulting in stronger density gradients transverse to the field. Consequently, radio waves are preferentially scattered in directions tangential to the local magnetic field. Similarly, high anisotropy of the density inhomogeneity in the interstellar medium has been inferred from the observations of ISS, which is regulated by the interstellar magnetic fields \citep{2010ApJ...708..232B,2014MNRAS.442.3338P,2014MNRAS.440L..36P}. Anisotropy in the density inhomogeneity of the solar corona have also been long studied (e.g. \cite{2025arXiv250601632Z}). Similar models of plasma screens composed of field-aligned columns of enhanced or depleted plasma density were also proposed to explain the modulation lanes in Jovian radio emission \citep{1997JGR...102.7127I,2013Icar..226.1214A}.

In addition, to have an appreciable amount of amplification, the width of the AFLs (at a single frequency) along the $x$-axis has to be smaller than the refractive scale $r_{\rm ref}$ of the plasma screen. For the assumed plasma screen with $\zeta\sim 200\;\rm rad$, we estimate that the radio-emitting AFLs has to be thinner than $\sim 600\;\rm km$ at a single frequency. This suggests that the elementary emitters of the S-bursts are spatially clustered over the typical lifetime of the modulation lanes ($\approx1\;\rm s$) at a single frequency.

\subsection{S-burst striae}

The S-burst striae might be interpreted as another form of intensity modulation superimposed on the original S-bursts. In ISS, short-duration and narrowband emission variations, often referred to as ``scintle'', are usually attributed to diffractive scintillation. The interference between two scattered images produces a periodic fringe pattern in the dynamic spectra \citep{2004MNRAS.354...43W}. In pulsar scintillation, diffractive scattering has distinct properties in the secondary spectra, revealing structures like the parabolic arcs and inverted arclets which provide valuable diagnostics to the scattering medium \citep{2004MNRAS.354...43W,2006ApJ...637..346C}. We do not expect to find similar structures in our case because of the (instantaneously) narrowband nature of our emission and its distinctive frequency-to-position mapping. As such, we limit {our} discussion to the period ($\approx 15\;\rm ms$) and the bandwidth ($\approx10\; \rm MHz$) of the striae in the framework of diffractive scattering. 

It is straightforward to show that the time scale of the striae is inconsistent with diffractive scintillation from a close-in plasma screen that we have already postulated. The typical time scale of the fringes from diffractive scintillation is determined by the transverse crossing time over the diffractive scale $r_F/\zeta$. Assuming the same Fresnel scale of $r_F\approx 3\;\rm km$ and the same transverse velocity on the order of $v_t\sim10^4\;\rm km/s$, as $\zeta\gg1$ for a strong scattering case, we obtain a typical diffractive time scale much shorter than a millisecond. {Such short-timescale variability, if present, cannot be resolved with the current time resolution of our observations. A second plasma screen, preferably located further away and moving at a much slower velocity, might provide a possible explanation on the time scale; however, this model is highly unconstrained, and we therefore do not discuss this possibility further.} 

The decorrelation bandwidth of the ``scintles'' has a frequency dependence, ($\Delta\nu\propto\nu^{4.4}$ in ISS, \cite{1985ApJ...288..221C}), which is regarded as a hallmark of diffractive scattering. However, the analysis only applies to compact sources emitting simultaneously broadband emission. The emitters of S-bursts produce intrinsically narrowband emission ($\approx 50 \;\rm MHz$) which also shifts in frequency as the source moves to different locations. This poses a major challenge for further analysis within the current work, as the physical nature of the S-burst sources and their relationship with a possible diffractive screen remain largely unknown.

Other non-local plasma propagation effects could be generally excluded. Such short timescale variability is not seen in ISS and a strong interstellar plasma lens is extremely rare within 5 pc (with rare exceptions, such as \cite{2015A&A...574A.125D}). Interplanetary scintillation is effective only close to the Sun \citep{1978SSRv...21..411C}, while our observations were conducted at midnight. Ionospheric scintillation is not an important factor to consider at our observing band even during the strongest ionospheric disturbances. 

We do not exclude the possibility of an intrinsic explanation for the striae. For instance, striae are found in solar Type III bursts, sometimes referred to as Type IIIb bursts \citep{2018ApJ...856...73C,2021NatAs...5..796R,2023ApJ...946...33C}. These features are thought to be related to the scattering of Langmuir waves, the trigger of plasma emission, by local plasma turbulence\citep{2021NatAs...5..796R}. The S-bursts from AD Leo come from a different emission mechanism and a direct comparison is not meaningful for interpretation. Morphologically, the striae observed in AD Leo S-bursts seem to be more organized in the dynamic spectra than the solar ones, which are mostly stochastic. Furthermore, the S-bursts have distinctive properties in the secondary spectra which have not been detected or explored in solar observations. Any intrinsic mechanism proposed to explain the AD Leo striae must account for the following: (1) some S-bursts have striae while some others do not; (2) the striae have a internal period an order of magnitude shorter than the period of the S-bursts; (3) striae have a similar but different drift rate compared to the S-bursts; and (4) striae introduce highly unusual patterns in the secondary spectra. 

\section{Summary and discussion}

We present the high time-resolution radio dynamic spectra of AD Leo on Dec. 1st, 2023. Over 15 min observations, we identified three distinct types of spectro-temporal structures, two of which, the modulation lanes and the striae, have not been reported before. Modulation lanes are broadband, second-long structures that show a general drift to lower frequencies. In most cases, they appear randomly while over a certain time period, a 1\;s quasi-periodicity is noticed. There are two types of sub-second scale structures in the data, the S-burst envelopes and S-burst striae. The two emission structures are periodic, with $\approx 0.1\,$s for the S-burst envelopes and $\approx0.01\,$s for the striae. Both S-burst envelopes and striae have positive frequency drifts, with the slope of striae being slightly higher. These three spectro-temporal structures are superimposed to one another in the dynamic spectra. We used DFTs and ACFs to separate the S-burst envelopes and striae and found very peculiar patterns in the secondary spectra. The secondary spectra reveal an intricate convolution of different emission patterns.

The identification of S-bursts alongside two additional patterns confirms that millisecond-scale bursts are ubiquitous in the radio emission from AD Leo, though they can exhibit very diverse morphologies.
We attribute the modulation lanes to refraction caused by inhomogeneous, regularly structured plasma in AD Leo's magnetosphere that crosses the LOS. Specifically, we propose that the modulation lanes arise as the radio source crosses the caustic points of the plasma screen. By modelling an idealised sinusoidal plasma phase screen, we are able to explain the magnification, shape, and quasi-periodicity of the modulation lanes. The spectro-temporal characteristics of the modulation lanes impose joint constraints on the distance, transverse velocity and density fluctuation of the scattering medium. We fail to reach a satisfying explanation for the spectral and temporal scales of the striae, which might also be a propagation effect or an intrinsic emission mechanism that is currently not understood.

The propagation effects we propose to be active in the magnetosphere of AD Leo differ significantly from those typically observed in the solar corona. On the Sun, scattering of radio emission is generally attributed to stochastic turbulence in the coronal plasma, occurring over a significant extent of the LOS \citep{2017NatCo...8.1515K,2019ApJ...884..122K}. Observable manifestations of such scattering include angular broadening of intrinsically compact sources \citep{2017NatCo...8.1515K,2023ApJ...956..112K}, displacement of the apparent position \citep{2017NatCo...8.1515K,2018ApJ...868...79C,2021ApJ...909..195Z}, temporal delay and broadening of the bursts \citep{2023MNRAS.520.3117C,2025ApJ...978...73C}, and certain directivity pattern of the emission \citep{2008A&A...489..419B,2021A&A...656A..34M}. In the solar case, there is no evidence for a localized, thin plasma layer that dominates the scattering process, nor are coherent time-frequency structures similar to those as we observe are reported and attributed to scattering. We believe that our results show more resemblance with propagation effects seen in Jovian radio emission due to refraction from an inhomogeneous plasma torus region \citep{1970A&A.....4..180R,1978Ap&SS..56..503R,1997JGR...102.7127I,2009A&A...493..651L,2013Icar..226.1214A}. The physical nature of the locally overdense scattering medium in AD Leo's magnetosphere is unconstrained. It might be analogous to giant coronal loops found on some chromospheric-active stars that extend to a few stellar radii \citep{2010Natur.463..207P,2020A&A...641A..90C}.

Scintillations studies offer great diagnostic potential, not only for probing the magnetospheric environment around stars, but also for determining the source size of the radio emitters. Two key ingredients are required for scintillation to occur: (1) a compact source that is smaller than the relevant refractive or diffractive scale (2) a regularly-structured thin plasma region capable of inducing strong scattering. We propose that similar phenomena might be found on other stars as well, especially those which are radio-active and have an extended magnetosphere. Based on this work, high time resolution radio observations (on seconds or sub-seconds) seem to be a requirement to carry out such studies. Apart from encouraging further studies using existing telescopes, we look forward to the next-generation radio telescopes, such as the FAST core array\citep{2024AstTI...1...84J}, the Square Kilometre Array (SKA, \cite{2009IEEEP..97.1482D}, and the next generation Very Large Array (ngVLA, \cite{2019clrp.2020...32D}), to provide more sensitive and higher resolution observations to help shed light on this emerging field.

\section*{acknowledgement} 
H.T. \& J.Z. acknowledge funding by the National Natural Science Foundation of China (grants 12425301 \& 12250006) and the Specialized Research Fund for State Key Laboratory of Solar Activity and Space Weather. J.Z. also acknowledges the support from the China Scholarship Council (No. 202306010244). HKV acknowledges funding from the Dutch research council's (NWO) talent programme (Vidi grant VI.Vidi.203.093) and from the European Research Council's starting grant `Stormchaser' (grant number 101042416). JRC acknowledges funding from the European Union via the European Research Council (ERC) grant Epaphus (project number 101166008).  This work made use of the data from FAST (Five-hundred-meter Aperture Spherical radio Telescope). FAST is a Chinese national mega-science facility, operated by National Astronomical Observatories, Chinese Academy of Sciences.

%

\vspace{5mm}
\facilities{FAST}


\software{astropy \citep{2013A&A...558A..33A,2018AJ....156..123A} Matplotlib \citep{2007CSE.....9...90H}, NumPy \citep{2020Natur.585..357H}, SciPy\citep{2020NatMe..17..261V}}



\appendix

\section{The low-pass filter for image processing}\label{sec:low-pass}

We used a 2D Butterworth filter, a type of low-pass filter, to remove the high frequency (small-scale) structures in the dynamic spectra while preserving the low frequency (large-scale) components. The Butterworth filter introduces a smooth gain function which transits from 1 to 0 near the cut-off frequency and has the advantage of avoiding ripple artefacts in the image.

Let $I(t,\nu)$ be the intensity map of the original dynamic spectra and $\Tilde{I}(f_t,f_\nu)$ be its Fourier conjugate: 
\begin{equation}
    \Tilde{I}(f_t,f_\nu)=\mathrm{DFT}(I(t,\nu)).
\end{equation}

We define the gain function of the Butterworth filter in the Fourier domain as 
\begin{equation}
    G(f_t,f_\nu)=\frac{1}{1+\left(f_t^2/f_{t_0}^2+f_\nu^2/f_{\nu_0}^2\right)^n},
\end{equation}

where $f_{t_0}$ and $f_{\nu_0}$ are the cut-off frequencies on the two Fourier axes and $n$ is the order {of the filter}. The contour lines of $G(f_t,f_\nu)$ are elliptical. We define the shape of power kernel of $G(f_t,f_\nu)$ as
\begin{equation}
    f_t^2/f_{t_0}^2+f_\nu^2/f_{\nu_0}^2=1,
\end{equation}

where the power gain of $G(f_t,f_\nu)$ drops to one half of the origin. The smoothed dynamic spectrum is obtained from inverse Fourier transform of the product between $\Tilde{I}(f_t,f_\nu)$ and $G(f_t,f_\nu)$
\begin{equation}
    I'(t,\nu)=\mathrm{DFT^{-1}}[\Tilde{I}(f_t,f_\nu)G(f_t,f_\nu)].
\end{equation}

$f_{t_0}=30\,$Hz, $f_{\nu_0}=0.05\,\mu$s, $n=3$ are adopted for image processing throughout the paper. The kernel of $G(f_t,f_\nu)$ has been shown as a red dashed ellipse in Figure \ref{fig:figure5}(a).

\section{A generic mathematical model of drifting bursts and secondary spectra}\label{sec:topology}

We discuss here the characteristic secondary spectra corresponding to a collection of uniform drifting bursts. DFT is a discrete analogue of continuous Fourier transform. Their relationship is described by
\begin{equation}
    \begin{split}
        \mathrm{DFT}[I(t,\nu)]&=\sum_{(m,n)}I(t_m,\nu_n)\exp \left[-2\pi i (f_t t_m+f_\nu\nu_n)\right]\\
        &=\iint I(t,\nu)\exp \left[-2\pi i (f_t t+f_\nu\nu)\right]\sum_{(n,m)}\delta(t-t_m)\delta(\nu-\nu_n) dtd\nu.
    \end{split}
\end{equation}

The subscripts $n,m$ are the pixel indexes and the term $\sum_{(n,m)}\delta(t-t_m)\delta(\nu-\nu_n)$ is the sampling function on a finite, regular grid. To understand the properties of the DFT of the dynamic spectra, we approximated the analytical solution using continuous Fourier transform for simplicity in the following analysis. There are two discrepancies which should be noted.

First, DFT is applied to finite-length signal segments, which is mathematically equivalent to multiplying an infinite-duration signal by a rectangular window (2D in our case). In the Fourier domain, this corresponds to a convolution with a a (2D) sinc function, the Fourier transform of the rectangular window, leading to the so-called spectral leakage. This leakage results in a cross-shaped power dispersion in the secondary spectra, which is not clearly visible in our observations probably due to limited dynamic range. Additionally, the periodic, discrete nature of the sampling causes the power pattern to be periodically replicated in the Fourier domain. The spacing between each replica is equal to the inverse of the resolution of the original data. This is generally avoided when employing Fast Fourier Transform, the fast algorithm of DFT, where computation is performed within one half of the inverse resolution.

We first consider the intensity of a single burst following a 2D Gaussian function, with time $t$ and frequency $\nu$ in the unit of pixel number,
\begin{equation}
    I(t,\nu)=I_0\exp{\left[-\frac{(t-t_0)^2}{2\sigma_t^2}-\frac{(\nu-\nu_0)^2}{2\sigma_\nu^2}\right]}.
\end{equation}

$(t_0,\nu_0)$ is the centroid of the burst and $(\sigma_t,\sigma_\nu)$ are the Gaussian widths. The half-maximum widths of the burst in the time and frequency axis are given by
\begin{equation}
    w_t=2\sqrt{2\ln2}\sigma_t,\quad w_\nu=2\sqrt{2\ln2}\sigma_\nu.
\end{equation}

The burst has the shape of an ellipse, with major axis along $t$ axis for $\sigma_t>\sigma_\nu$. It can be proven that the Fourier conjugate of a 2D Gaussian function is also in the form of a 2D Gaussian function,
\begin{equation}
    \Tilde{I}(f_t,f_\nu)=\Tilde{I}_0\exp\left[-\frac{f_t^2}{2\sigma_{f_t}^2}-\frac{f_\nu^2}{2\sigma_{f_\nu}^2}\right]\exp\left[-2\pi i(f_tt_0+f_\nu\nu_0)\right],
\end{equation}
\begin{equation}
    \sigma_{f_t}=\frac{1}{2\pi\sigma_t},\quad \sigma_{f_\nu}=\frac{1}{2\pi\sigma_\nu}, \quad \Tilde{I}_0=2\pi\sigma_t\sigma_\nu I_0.
\end{equation}

The power pattern in the Fourier domain is also elliptical, with the major axis along $f_\nu$ axis for $\sigma_t>\sigma_\nu$. The phase term of $\Tilde{I}(f_t,f_\nu)$ is related to the burst centroid position on the $(t,\nu)$ plane.

For a group of uniform elliptical bursts, the Fourier conjugate is the sum of their complex contributions,
\begin{equation}
    \Tilde{I}(f_t,f_\nu)=\Tilde{I}_0\exp\left[-\frac{f_t^2}{2\sigma_{f_t}^2}-\frac{f_\nu^2}{2\sigma_{f_\nu}^2}\right]\sum_j\exp\left[-2\pi i(f_tt_j+f_\nu\nu_j)\right],
\end{equation}

where $j$ is the index of each burst and $(t_j,\nu_j)$ is the centroid position of the burst.

We modelled a frequency-drifting burst as a similar 2D Gaussian function, rotated counter-clockwise with an angle of $\theta$. We define the major and minor axes of such an oblique burst as $t'$ and $\nu'$ axes. The coordinate transformation follows
\begin{equation}
    (\boldsymbol{r}-\boldsymbol{r_0})^\mathsf{T}=\boldsymbol{R}\boldsymbol{r'}^\mathsf{T},
\end{equation}
\begin{equation}
    \boldsymbol{r}=(t,\nu),\quad \boldsymbol{r_0}=(t_0,\nu_0),\quad\boldsymbol{r'}=(t',\nu'),
\end{equation}
\begin{equation}
    \boldsymbol{R}=\left[
    \begin{array}{cc}
    \cos\theta     &-\sin\theta  \\
    \sin\theta    & \cos\theta
    \end{array}\right],
\end{equation}

where $\boldsymbol{R}$ is the rotation matrix and $\mathsf{T}$ is the transpose operator. The intensity distribution function is given by
\begin{equation}
    I'(t,\nu)=I(t',\nu')=I_0'\exp\left[-a(t-t_0)^2-2b(t-t_0)(\nu-\nu_0)-c(\nu-\nu_0)^2\right]
\end{equation}
\begin{equation}
    a=\frac{\cos^2\theta}{2\sigma_{t'}^2}+\frac{\sin^2\theta}{2\sigma_{\nu'}^2},\quad b=\frac{\sin\theta\cos\theta}{2\sigma_{t'}^2}-\frac{\sin\theta\cos\theta}{2\sigma_{\nu'}^2},\quad c=\frac{\sin^2\theta}{2\sigma_{t'}^2}+\frac{\cos^2\theta}{2\sigma_{\nu'}^2}.
\end{equation}

When $\sigma_{t'}>\sigma_{\nu'}$, the major axis of such an oblique burst is described by
\begin{equation}
    (t-t_0)\sin\theta-(\nu-\nu_0)\cos\theta=0
\end{equation}

The slope of such a burst is $k_1=\tan\theta$. The Fourier conjugate of $I'(t,\nu)$ is given by
\begin{equation}
    \Tilde{I}'(f_t,f_\nu)=\Tilde{I}'_0\exp{\left(-\alpha f_t^2-2\beta f_t f_\nu-\gamma f_\nu^2\right)}\exp\left[-2\pi i(f_tt_0+f_\nu\nu_0)\right],
\end{equation}
\begin{equation}
    \alpha=\frac{\cos^2\theta}{2\sigma_{f_{t'}}^2}+\frac{\sin^2\theta}{2\sigma_{f_{\nu'}}^2},\quad \beta=\frac{\sin\theta\cos\theta}{2\sigma_{f_{t'}}^2}-\frac{\sin\theta\cos\theta}{2\sigma_{f_{\nu'}}^2},\quad \gamma=\frac{\sin^2\theta}{2\sigma_{f_{t'}}^2}+\frac{\cos^2\theta}{2\sigma_{f_{\nu'}}^2},
\end{equation}
\begin{equation}
    \sigma_{f_{t'}}=\frac{1}{2\pi\sigma_{t'}},\quad \sigma_{f_{\nu'}}=\frac{1}{2\pi\sigma_{\nu'}}, \quad \Tilde{I}_0'=2\pi\sigma_{t'}\sigma_{\nu'} I_0'
\end{equation}

The absolute value of the $\Tilde{I}'(f_t,f_\nu)$ is a rotation of that of $\Tilde{I}(f_t,f_\nu)$ in equation (B8), in accord with the same rotation matrix $\boldsymbol{R}$. The slope of the major axis in the Fourier domain is $k_2=-1/\tan\theta$, perpendicular to the slope of the burst. Similarly, we have the Fourier conjugate for a group of bursts in uniform shape
\begin{equation}
    \Tilde{I}'(f_t,f_\nu)=\Tilde{I}'_0\exp{\left(-\alpha f_t^2-2\beta f_t f_\nu-\gamma f_\nu^2\right)}\sum_j\exp\left[-2\pi i(f_tt_j+f_\nu\nu_j)\right],
\end{equation}

And the secondary spectrum is given by
\begin{equation}
    \Tilde{S}'(f_t,f_\nu)=\Tilde{I}'^2_0 \exp{\left(-2\alpha f_t^2-4\beta f_t f_\nu-2\gamma f_\nu^2\right)}\sum_{j,k}\cos\left[2\pi (f_t(t_j-t_k)+f_\nu(\nu_j-\nu_k))\right],
\end{equation}

where $j,k$ are the indexes of the bursts ($j,k$ could be the same). From equation (B21), we find that the power in the secondary spectrum of a group of bursts has two components. One stems from the Fourier conjugate of the intensity distribution function of a single burst, whose power is a 2D Gaussian function. The other is related to the positions of the bursts, or more precisely, the Delta-function of the centroids of the bursts $\sum_j\delta(t-t_j)\delta(\nu-\nu_j)$. If the bursts are randomly distributed in the time-frequency plane, the first component dominates, manifested as an elliptical shape of power at the center of the secondary spectra. If the occurrence of the bursts is very regular, for instance, equally-spaced in time, strong power at $f_t=1/\Delta t$ (where $\Delta t$ is the interval between two adjacent bursts) and its harmonics should be observed. 

We would like to re-write the intensity distribution function equation (B14) in the form of
\begin{equation}
    I'(t,\nu)=I'_0\exp{\left\{-\frac{[\nu-\nu_0-k(t-t_0)]^2}{2\sigma_\nu^2}-\frac{(t-t_0)^2}{2\sigma_t^2}\right\}},
\end{equation}
\begin{equation}
    k=-\frac{b}{c},\quad \sigma_\nu=\frac{1}{\sqrt{2c}},\quad\sigma_t=\frac{1}{\sqrt{2a-\frac{2b^2}{c}}}.
\end{equation}

We define $\nu-\nu_0-k(t-t_0)=0$ as the spline of the burst, which is the trajectory of the frequency with maximum intensity at each time. The spline approximately coincides with the major axis of the elliptical burst, as when $\sigma_{t'}\gg\sigma_{\nu'}$, $-b/c\approx\tan\theta$. 

We also simulated the case of a curved burst, with the slope varying with frequencies. The intensity distribution function is given by
\begin{equation}
    I''(t,\nu)=I''_0\exp{\left\{-\frac{[\nu-\nu_0-k(t-t_0)-\eta(t-t_0)^2]^2}{2\sigma_\nu^2}-\frac{(t-t_0)^2}{2\sigma_t^2}\right\}}.
\end{equation}

The spline of the burst is 
\begin{equation}
    \nu-\nu_0-k(t-t_0)-\eta(t-t_0)^2=0.
\end{equation}

The changing drift rate is
\begin{equation}
    \frac{d\nu}{dt}=k+2\eta(t-t_0)\approx k+\frac{2\eta}{k}(\nu-\nu_0).
\end{equation}

The approximation holds when the curvature of the burst $\eta$ is relatively small, $\eta(t-t_0)\ll k$. It is hard to derive an analytical solution of the power of its Fourier conjugate and we used simulation-based approach to assess the basic properties. In a simplified view, the curved burst can be approximated as a series of linear bursts, each with a slightly different slope. As a result, the corresponding secondary spectrum can be interpreted as the superposition of multiple oblique ellipses, reflecting the dispersion in the slopes.

\begin{figure*}[htb!]
    \centering
    \includegraphics[width=0.8\linewidth]{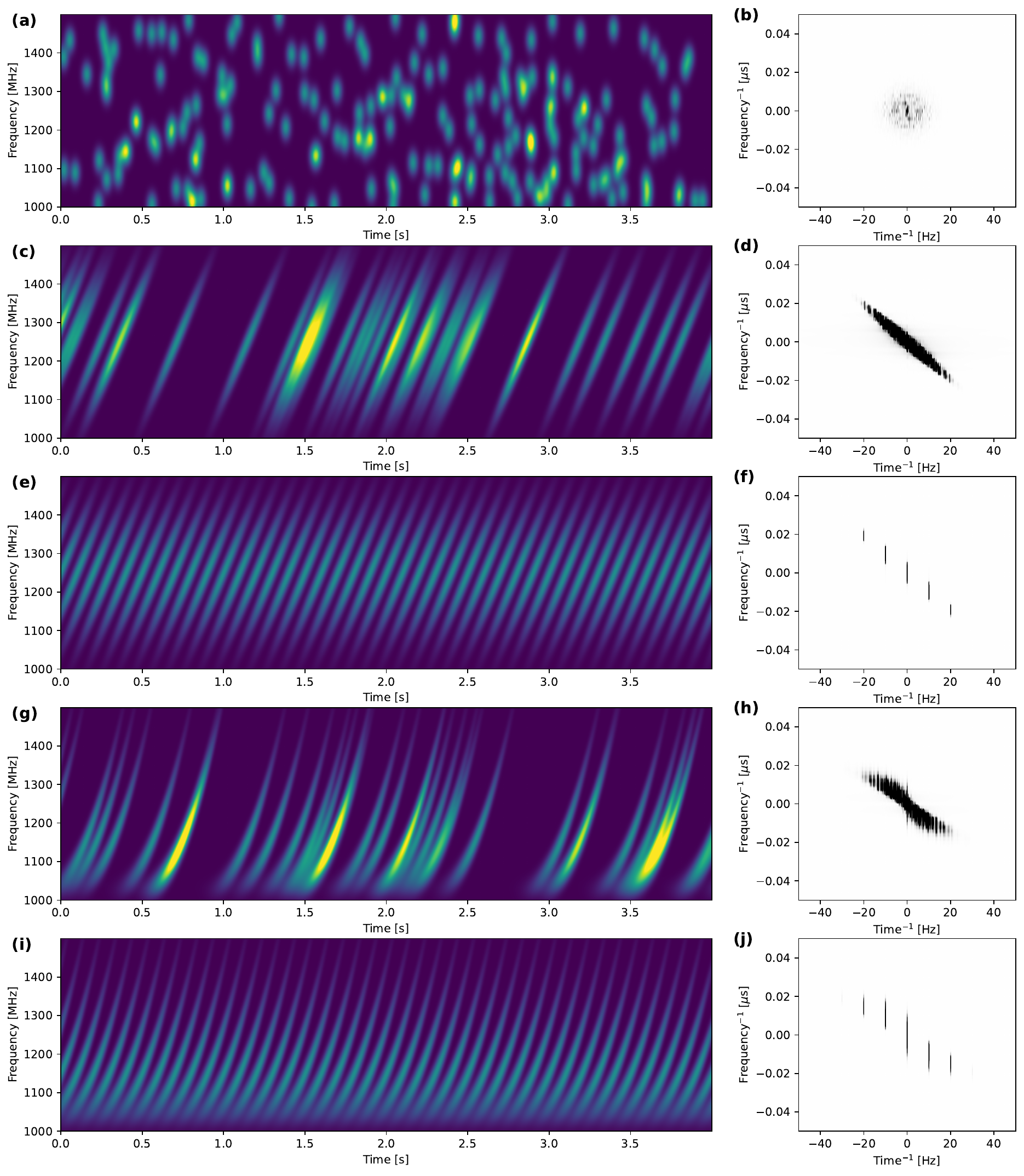}
    \caption{Simulation of dynamic spectra and secondary spectra for a set of uniformly shaped bursts. The colour scale in the dynamic spectra is the intensity from 0 to 2 and the grey scale in the secondary spectra is the SNR from 0 to $10^7$. Both are shown on a linear scale. (a,b) Random blob bursts. (c,d) Random stripe bursts. (e,f) Periodic stripe bursts. (g,h) Random curved bursts. (i,j) Periodic curved bursts.}
    \label{fig:figure10}
\end{figure*}

We simulated the dynamic spectra with a bandpass of 1.0 - 1.5 GHz and a duration of 4 s. The time and frequency resolutions were set at 0.4 ms and 0.5 MHz, consistent with the resolutions of the observed dynamic spectra. We considered five cases which could be attributed to the intrinsic collection of the bursts (Figure \ref{fig:figure10}). Equation (B24) was used as the general form of intensity distribution of the bursts.

(1) \textit{Random blob bursts} (Figure \ref{fig:figure10}(a,b)). $k=\eta=0$, $\sigma_t=20\;$ms, $\sigma_\nu=20\;$MHz. $\nu_0$ and $t_0$ are random in the time-frequency plane. The corresponding secondary spectrum has a blob of strong power in the center following a 2D Gaussian function.

(2) \textit{Random stripe bursts} (Figure \ref{fig:figure10}(c,d)). $\eta=0$, $k=1000\;$MHz/s, $\sigma_t=100\;$ms, $\sigma_\nu=20\;$MHz, $\nu_0=1250\;$MHz. $t_0$ are random in time. The corresponding secondary spectrum is a tilted ellipse, with the slope being perpendicular to the slopes of the bursts in the dynamic spectrum.

(3) \textit{Periodic stripe bursts} (Figure \ref{fig:figure10}(e,f)). $\eta=0$, $k=1000\;$MHz/s, $\sigma_t=100\;$ms, $\sigma_\nu=20\;$MHz, $\nu_0=1250\;$MHz. The values of $t_0$ are periodic in time, with a period of 0.1 s. As a result, the secondary spectrum exhibits strong power at $f_t=10\;$Hz and its harmonics. The power pattern also has an elliptical envelope similar to Figure \ref{fig:figure10}(d).

(4) \textit{Random curved bursts} (Figure \ref{fig:figure10}(g,h)). $\eta=2000\;$MHz/s$^2$, $k=1000\;$MHz/s, $\sigma_t=100\;$ms, $\sigma_\nu=20\;$MHz, $\nu_0=1150\;$MHz. $t_0$ are random in time. The secondary spectrum is similar to Figure \ref{fig:figure10}(d), except for the dispersion in the slopes.

(5) \textit{Periodic curved bursts} (Figure \ref{fig:figure10}(i,j)). $\eta=2000\;$MHz/s$^2$, $k=1000\;$MHz/s, $\sigma_t=100\;$ms, $\sigma_\nu=20\;$MHz, $\nu_0=1150\;$MHz. The values of $t_0$ are periodic in time, with a period of 0.1 s. The secondary spectrum exhibits strong power at $f_t=10\;$Hz and its harmonics. The power pattern has an envelope in the shape of Figure \ref{fig:figure10}(h).

It seems that this simple model considering only uniform drifting bursts fails to reflect the diverse properties that are seen in the observed secondary spectra. As a heuristic approach, we consider a multiplicative dynamic spectrum formed by the product of two independent patterns. The first is a set of randomly distributed stripe-like bursts, as shown in Figure \ref{fig:figure10}(c). The second is a periodic fringe pattern, described by a cosine function,
\begin{equation}
    I'''(\nu,t)=1+m\cos(2\pi f_\nu'(\nu-k't)).
\end{equation}

\begin{figure*}[htb!]
    \centering
    \includegraphics[width=0.8\linewidth]{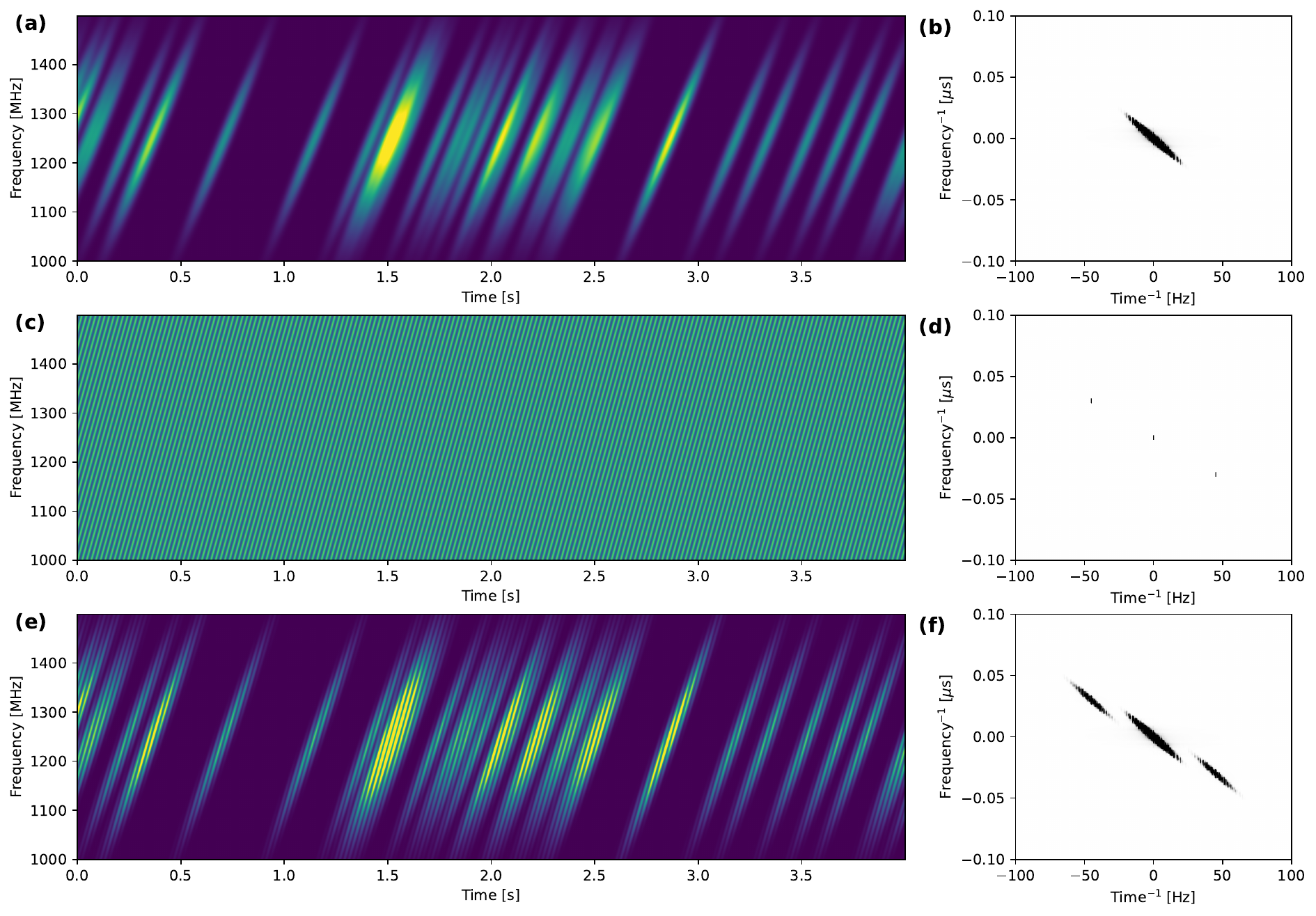}
    \caption{Simulation of a multiplicative pattern in the dynamic spectrum and the corresponding secondary spectrum. The color scale and the grey scale are the same with Figure \ref{fig:figure10}. (a,b) Random stripe bursts. (c,d) Periodic fringe pattern. (e,f) Multiplicative dynamic spectrum and convolved secondary spectrum.}
    \label{fig:figure11}
\end{figure*}

The multiplication in the time-frequency plane corresponds to a convolution in the Fourier domain. We set $m=0.5$, $f_\nu'=0.03\;\mu$s, $k'=1500\;$MHz/s for the periodic fringe pattern and show the results as below. The convolved secondary spectra (Figure \ref{fig:figure11}(e)) is topologically similar to the observed one in Figure \ref{fig:figure5}(d).

\section{Measuring the spectro-temporal properties of the S-burst envelopes and striae}\label{sec:drift}

\begin{figure*}[htb!]
    \centering
    \includegraphics[width=0.9\linewidth]{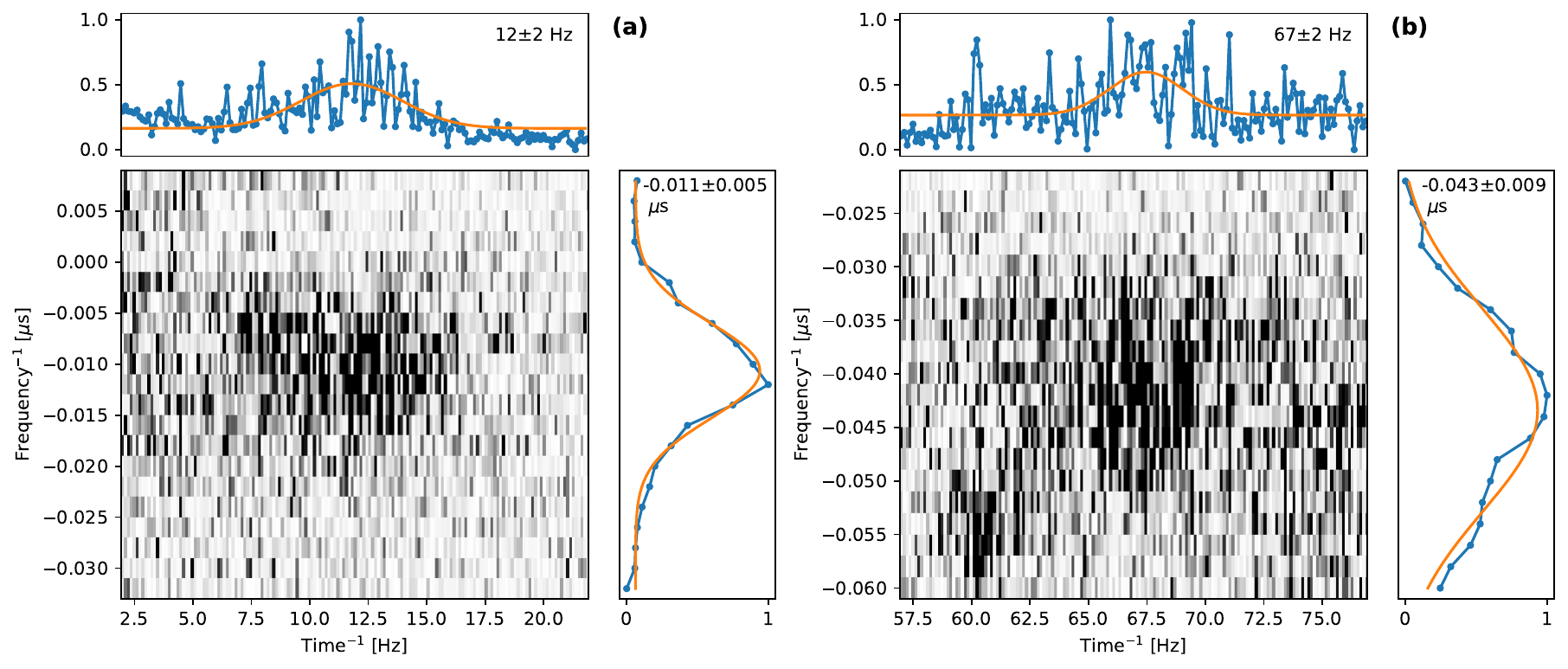}
    \caption{The fit of the centroid locations of the inner knot (a) and outer knot (b) in the secondary spectrum of Figure \ref{fig:figure5}(a). The blue curve in the top (right) sub-panel is the normalized power integrated along the vertical (horizontal) axis. The orange curve shows the Gaussian fit, with the peak and one-sigma width indicated.}
    \label{fig:figure12}
\end{figure*}

We used the secondary spectra and ACFs to characterize the properties of the S-burst envelopes and striae, as summarized in Table \ref{tab:table1}. The parameters include periodicity, slope, instantaneous bandwidth, and single-frequency duration. The periodicity of the S-burst envelopes (striae) was determined by the location of the inner (outer) knot in the secondary spectra. As illustrated in Figure \ref{fig:figure12} as an example, we selected a region centered on a knot and integrated the power along both frequency axes in the Fourier domain. The resulting profiles were then fitted with a Gaussian function, with the peak and one-sigma width of the fit taken as the measured value and its associated uncertainty.

We approximated the instantaneous bandwidth $w_\nu$ and single-frequency duration $w_t$ of the structure (S-burst envelope or striae) with
\begin{equation}
    w_\nu\approx\left|\frac{1}{2f_\nu}\right|,\quad w_t\approx\left|\frac{1}{2f_t}\right|.
\end{equation}

The slope could be derived by
\begin{equation}
    k\approx-\frac{f_t}{f_\nu}.
\end{equation}

However, this method tends to overestimate the uncertainty of the slope. This is because that, though $f_t$ and $f_\nu$ have large respective uncertainties, they are also correlated. To obtain a more robust estimate, we instead analyzed the fringe patterns in the ACFs.

\begin{figure*}[htb!]
    \centering
    \includegraphics[width=0.9\linewidth]{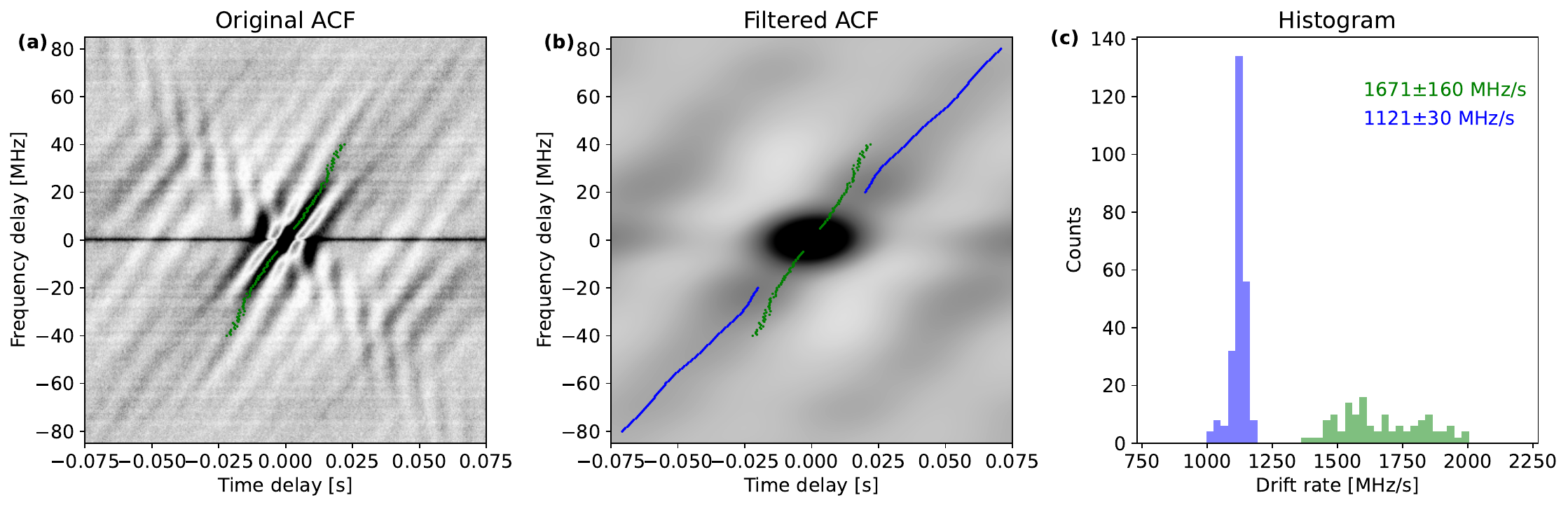}
    \caption{Measurement of the slopes of S-burst envelopes and striae using ACFs. (a) The trajectory of the central fringe in the original ACF of Figure \ref{fig:figure5}(b) shown as green scattered dots. (b) The same trajectory for the filtered ACF of Figure \ref{fig:figure5}(c) in blue scattered dots. (c) The slope distributions of the two trajectories. The blue distribution corresponds to the S-burst envelope and the green one represents the striae.}
    \label{fig:figure13}
\end{figure*}

The slope of the fringes in the ACF is a weighted average of the drifting slopes in the dynamic spectrum. We traced the trajectories of the central fringes in the ACFs of the original spectra (Figure \ref{fig:figure5}(b, e, h)) and the filtered spectra (Figure \ref{fig:figure5}(c, f, i, l)). An example is shown in Figure \ref{fig:figure13}. 

At each frequency delay, we selected the time delay corresponding to the maximum correlation power. To avoid contamination from adjacent fringes, we manually masked the nearby pixels. For the original ACF, we traced the trajectory from $|f_\nu|=5\;$MHz to 40 MHz and for the filtered ACF, we traced it from $|f_\nu|=20\;$MHz to 80 MHz. This is mainly to avoid the artifacts from the flagged frequency channels and the low-pass filter function. We obtained the slope distribution of the scattered dots and used the mean and the standard deviation as the measured slope and its uncertainty. The results of 1671$\pm$160 MHz/s for the stria and 1121$\pm$30 MHz/s for the S-burst envelopes generally agree with those obtained using equation (C29): 1558$\pm$329 MHz/s for the striae and 1091$\pm$528 MHz/s for the S-bursts (the uncertainties here are probably overestimated as the covariance between $f_t$ and $f_\nu$ is not accounted for).

\bibliography{sample631}{}
\bibliographystyle{aasjournal}



\end{document}